# Large-Area Conductor-Loaded PDMS Dielectric Composites for High-Sensitivity Wireless and Chipless Electromagnetic Temperature Sensors

Benjamin King, Nikolas Bruce, and Mahmoud Wagih*

University of Glasgow, James Watt School of Engineering, Glasgow, G12 8QQ, UK

*Mahmoud.Wagih@glasgow.ac.uk

## Abstract

Wireless electromagnetic sensing is a passive, non-destructive technique for measuring physical and chemical changes through permittivity or conductivity changes. However, the readout is limited by the sensitivity of the materials, often requiring complex sampling, particularly in the GHz range. We report capacitive dielectric temperature sensors based on polydimethylsiloxne (PDMS) loaded with 10 vol% of inexpensive, commercially available conductive fillers including copper powder (Cu), graphite powder (GP) and milled carbon fibre powder (CF). The sensors are tested in the range of 20°C – 110°C, with enhanced sensitivity from 20 - 60°C, and relative response of up to 85.5% at 200 MHz for CF-loaded capacitors. Additionally, we demonstrate that operating frequency influences the relative sensing response by as much as 15.0% in loaded composite capacitors. Finally, we demonstrate the suitability of PDMS-CF capacitors as a sensing element in wirelessly coupled chipless resonant coils tuned to 6.78 MHz with a readout response, in the resonant frequency of the sensor. The wireless readout of the PDMS-CF chipless system exhibited an average sensitivity of 0.38 %·°C$^{-1}$ which is a 40x improvement over a pristine PDMS-based capacitive sensor and outperforms state of the art frequency-domain radio frequency (RF) temperature sensors including carbon-based composites at higher loadings. Exploiting the high sensitivity, we interrogate the sensor wirelessly using a low-cost and portable open-source NanoVNA demonstrating a relative response in the resonant frequency of the reader coil of 48.5%, with the response in good agreement with the instrumentation-grade vector network analyzers (VNAs), demonstrating that the sensors could be integrated into an inexpensive and portable measurement setup that does not rely on specialty equipment or highly trained operators.

## 1. Introduction

With the rapid development of smart wearable devices for body area networks, there is greater demand for flexible, lightweight, and conformable large-area sensors which will enable more facile monitoring of physiological parameters.[1] Temperature is among the most fundamental measured physical parameters and reflects both the state of the object and the surrounding environment.[2–4] Measurement and control of temperature are important in many fields, including medicine, manufacturing and agriculture, further demonstrating the demand for the development of inexpensive, large-area sensors.

Functional polymer-based composites have demonstrated excellent potential as sensors primarily owing to the combination of properties of the polymer (low density, ease of processing, inherent flexibility and biocompatibility) and filler material (high thermal or electrical conductivity, excellent mechanical properties) being imparted to the resulting composite.[5] Composites consisting of a polymer matrix loaded with conductive filler in particular have been widely reported as smart flexible strain sensors and piezoelectric sensors based on the variation of electrical resistance due to external applied force,[6–8] resulting from interactions of adjacent conductive filler particles or fibers.[9] Conductive polymer composites have also recently been reported as temperature sensors, including poly(N-isopropylacrylamide-co-N-methylol acrylamide)/acid-treated carbon black which demonstrated a monomodal surface resistance trace resulting from moisture-induced swelling and deswelling of the composite with temperature.[10] Carbon nanomaterials including carbon nanotubes and graphene have also been widely studied in wearable electronics and temperature sensors owing to their excellent mechanical and chemical stability and high thermal conductivity.[11]

Sensors based on composites made from polydimethylsiloxane (PDMS) have been widely explored owing to the excellent  thermally stability, water and oxidation resistance, and biocompatibility of the polymer.[12,13] Applications of PDMS-loaded composites include stretchable conductive films for printed electronics,[14,15] and capacitive pressure sensors.[16–18] PDMS composites loaded with carbon nanotubes (CNTs) have shown some temperature-dependant response to strain, demonstrating that conductive PDMS composites have some temperature sensitivity.[19]

PDMS and conductive PDMS composites have been reported as temperature sensors, owing to their ability to undergo thermal expansion when heated.[20] Loading PDMS with conductive materials is one strategy for enhancing its temperature sensitivity,[21] since temperature fluctuation results in mechanical deformation of the polymer matrix, inducing separation of conductive networks and resulting in a change in the DC resistance or capacitance of the material (Figure 1A).[11] Conductive materials loaded into PDMS composites for enhancing its temperature sensitivity include graphite powder,[22,23] graphene,[24] carbon nanotubes,[25] and polyanaline.[26]

A primary limitation of PDMS composite temperature sensors is that they are read as direct current (DC) sensors, meaning that they can only extract sensing measurements by measuring the DC resistance of the material. Reading sensors with alternating current (AC) enables the complex impedance response to be read, and can deliver improved sensitivity and linearity of response at both kHz and GHz-frequency range.[27,28] This has been demonstrated for gas sensing[27] and temperature sensing using PDMS composites.[28] However, enhancements in temperature sensing have only been demonstrated in the GHz part of the electromagnetic spectrum, with no investigation of frequency-specific sensitivity enhanvement.[28] Additionally, a sensor with readout at GHz frequency, which detects sub-MHz or dB changes in the resonant frequency or amplitude, respectively, requires precisely calibrated vector network analysis (VNA) hardware,[29–36] limiting the practicality of these sensors and their ability to be read in varying and unknown environment outside of the laboratory, without a stable reference. Moreover, while passive chipless LC sensors have been explored for a variety of applications including biomedical sensors and in consumer products,[37–41] the low coupling between the reader and the sensor implies that specialized readout techniques are required,[42,43], with the observable resonance shift being limited to sub-MHz levels.

We present a low-cost wireless capacitive temperature sensor fabricated by loading a pure dielectric material (PDMS) with commercially-available conductive fillers (Figure 1B), investigating the readout sensitivity as a function of interfacing frequency, to enable a highly sensitive and portable wireless and chipless interface. These include graphite powder (90% < 44 μm diameter), milled carbon fiber (CF, 100 x 7.5 μm diameter), and copper powder (Cu, < 44 μm diameter) to fabricate a polymer matrix. The sensors were fabricated using a scalable moulding process and offer potential composite materials for the next generation of large-area electronics.

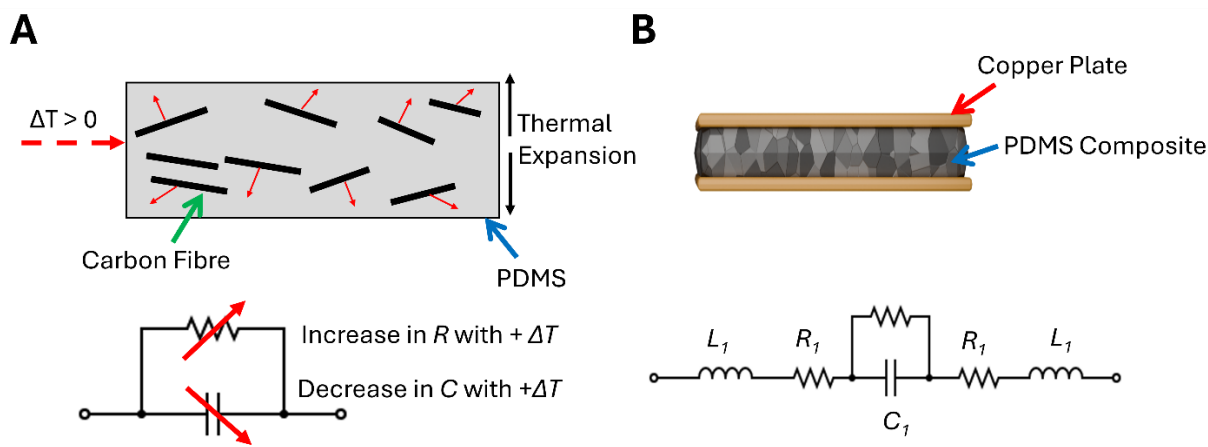

**Figure 1.** Capacitive sensing principle using conductor-loaded dielectric. A) Schematic of sensing mechanism of PDMS composite dielectric and equivalent circuit model of dielectric under temperature load. B) Schematic of PDMS composite capacitor and equivalent circuit model of capacitors in this work.[44]

## 2. Materials and Methods

### 2.1 Materials

Milled Carbon Fibre powder (D = 7.5 μm, L = 100 μm, FP-MCF-004), Graphite Powder (FP-GP-035) and Copper Metal Powder (325 mesh, FP-COP-025) were purchased from Easy Composites. SYLGARD™ 184 Silicone Elastomer Kit including polydimethylsiloxane (PDMS) and curing agent was purchased from Farnell.

### *2.2 Fabrication of loaded composite parallel plate capacitors*

A summary of the mass and volume fractions of filler in each loaded composite in this study is shown in Table 1. The requisite mass fraction of PDMS and filler is loaded into a beaker and stirred with a metal spatula until combined and well mixed. Next, the curing agent is added in a ratio of 1 part to 10 parts of the PDMS by volume and stirred with the metal spatula until combined and well mixed. The mixture is then stored in a glass desiccator connected to a vacuum pump and degassed for 30 minutes to remove air bubbles in the composite. Next, the mixture is cast into a glass petri dish (D = 90 mm), allowed to settle, and is degassed for an additional 10 minutes to remove remaining air bubbles. The degassed composites are then cured at 100 °C for 35 minutes for the pristine sample, and 100 °C for 1 hour for the composite samples. The cross-sectional morphology of the composites were characterized by scanning electron microscopy (SEM) images taken using a Hitachi SU8240 at voltage of 3.0 kV and a current of 5.0 μA.

The parallel plate capacitors were fabricated by cutting a 2 x 2 cm square of the composite and placing it between two pieces of conductive copper tape, as low-resistance electrodes.

The capacitor structure was then connected to standard Sub-miniature-A (SMA) RF Connector by soldering leads of < 1 cm to the signal and ground pins.

The capacitor samples were characterized using a Vector Network Analyzer (VNA) from Pico Technology (PicoVNA 106, covering 300 kHz – 6 GHz) connected to a phase-stable Mini-Circuits CBL-1.5M-SMSM+ cable (L = 1.5m) and a braided copper cable (L = 15 cm). The VNA was calibrated using a full two-port Short, Open, Load, Through (SOLT) traceable calibration kit (TA345 SOLT-STD-F). The broadband response of each device was measured up to 600 MHz between room temperature (20 °C) to 110 °C, by connecting devices to the VNA and securing it to a hot plate with polyimide tape (Figure S1, Supporting Information).

The capacitance (C) of parallel plate capacitors in this work is calculated using equation (1):

$$C = \frac{-1}{\text{Im}\{Z\} \cdot 2\pi f} \qquad \text{Im}\{Z\} < 0 \qquad (1)$$

Where $\text{Im}\{Z\}$ is the imaginary component of the measured complex impedance, and $f$ is the frequency. At lower frequencies, it can be assumed that the parasitic inductance is not significant below the self-resonant frequency where there is a transition between capacitance being dominant and inductance being dominant.[45]

**Table 1.** Composition of PDMS-filled composites fabricated for this study.

| Sample ID | Filler | Filler Mass Loading (wt %) | Filler Volume Loading (vol %) | Composite thickness (mm) |
|---|---|---|---|---|
| PDMS | None | 0 | 0 | 0.90 |
| PDMS-CF | Milled CF | 16.3 | 10 | 0.90 |
| PDMS-GP | Graphite Powder | 19.2 | 10 | 0.90 |
| PDMS-Cu | Cu Metal Powder | 49.1 | 10 | 0.90 |

### 2.3 Fabrication and characterization of wireless capacitor sensors

To achieve wireless sensing, we integrated temperature-sensitive capacitors with coupled inductive coils. The coils and resulting LC resonant circuits tuned to 6.78 MHz were designed using Keysight ADS resulting in a 9-turn, 10 x 10 cm square inductor (Figure S10) and fabricated on polyimide substrates with copper traces, in a standard flexible PCB process.

The resonant frequency ($f$) of the LC network changes in response to capacitance according to equation (2), where $L$ is the inductance and $C$ is the capacitance of the circuit.

$$f = \frac{1}{2\pi\sqrt{LC}}$$  (2)

The PDMS and PDMS-CF capacitors were soldered to stranded core wires and connected to an inductive coil in series to form the sensor. The reader coil was tuned to 6.78 MHz with a series capacitor of 66pF. The composite capacitor was placed on a hot plate identical to the one-port measurement setup and the coils were separated by 45 mm using a piece of Styrofoam. The entire system was then characterized in the range of 0.5 MHz to 30 MHz in the range of 20°C – 110°C using the standard VNA, or a portable NanoVNA,[46] in the sensing demonstration.

## 3. Results and Discussion

### 3.1 Capacitive properties of loaded PDMS composites

The parallel plate capacitors were characterized using the VNA up to 600 MHz to understand the influence of the composites' loading material on its dielectric properties (Figure 2). The PDMS-CF and PDMS-GP capacitors have a higher baseline capacitance and reduced quality factor compared to PDMS-Cu and pristine PDMS capacitors, which is attributed to the increased filler conductivity reducing the overall resistance in the lumped device. PDMS-Cu had a relatively low average capacitance and elevated Q approaching the behaviour of pristine PDMS devices, as opposed to the carbon conductor-loaded composites. This can be attributed to the oxidation of the copper powder after exposure to air post-manufacturing[47] and the resulting presence of a mixture of semiconducting copper oxide and small quantities of conductive copper. It is important to note that beyond 200 MHz, the calculated capacitance of all four composites begins to increase exponentially with frequency, as the devices approach their self-reosnant frequency.[45,48] To evaluate the losses in the loaded composites, the quality factor ($Q$) up to 600 MHz was calculated based on the ratio of the imaginary part of the complex impedance ($\mathrm{Im}\{Z\}$) and real part of the complex impedance ($\mathrm{Re}\{Z\}$) using:

$$Q = \left|\frac{\mathrm{Im}\{Z\}}{\mathrm{Re}\{Z\}}\right|.$$  (3)

The quality factor of the carbon-loaded composites PDMS-CF and PDMS-GP are six to eight-fold lower than PDMS up to 200 MHz attributed to the conductive (lossy) fillers in the composite, increasing its loss tangent ($\tan\delta = 1/Q$). The frequency-dependant quality factor of the PDMS-Cu capacitor follows a similar trend to that of pristine PDMS but is consistently lower in magnitude, owing to the Cu imparting some influence on the complex permittivity of the material and making the capacitor less ideal.[49] The self-resonant frequency of each composite, at which $\mathrm{Im}\{Z\} = 0\ \Omega$ is shown in Figure 2C. The carbon-loaded composites PDMS-CF and PDMS-GP have self resonant frequencies of 284.0 MHz and 300.5 MHz, respectively, which are significantly lower than, pristine PDMS. This is attributed to the inverse relationship between resonant frequency and capacitance outlined in equation 2, and the higher capacitance observed in Figure 2(a).

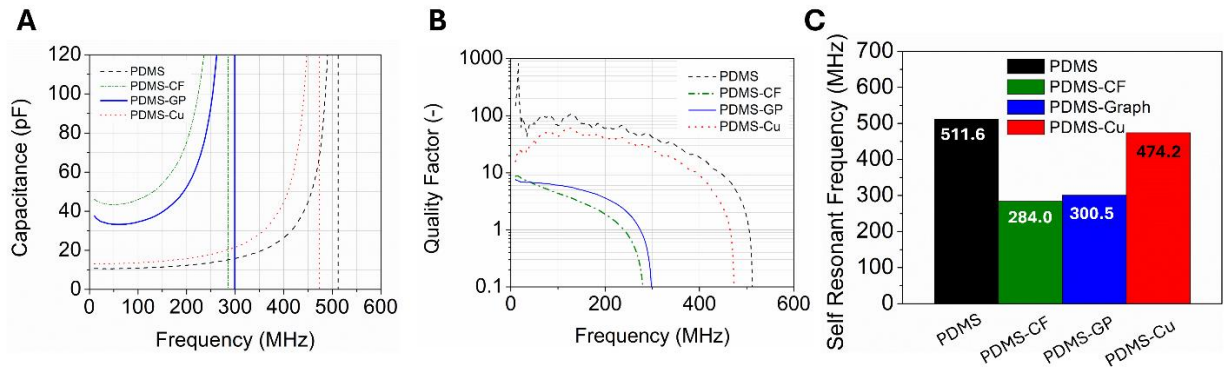

**Figure 2.** Room-temperature broadband composite-based capacitor electrical properties: A) capacitance as a function of frequency from 10 MHz to 600 MHz for composite-loaded capacitors, B) Quality factor Q at room temperature and C) Self-resonant frequency of each capacitor.

To understand the distribution of the loaded material in the polymer composite matrix and develop relationships between their morphologies and electrical properties (Figure 3), SEM micrographs were taken of the dielectric. PDMS-GP (Figure 3C and 3G) and PDMS-CF (Figure 3D and 3H) composites have uniformly distributed and predominantly randomly oriented conductive filler throughout the polymer, indicating that they are well mixed. However, PDMS-Cu (Figure 3B and 3F) composites have regions of high Cu powder concentration and low Cu concentration throughout the cross section of the composite, which is attributed to its higher density of 6.31 g ·cm³, compared to 1.8 g·cm³ of the milled CF and 2.6 g·cm³ of graphite powder.

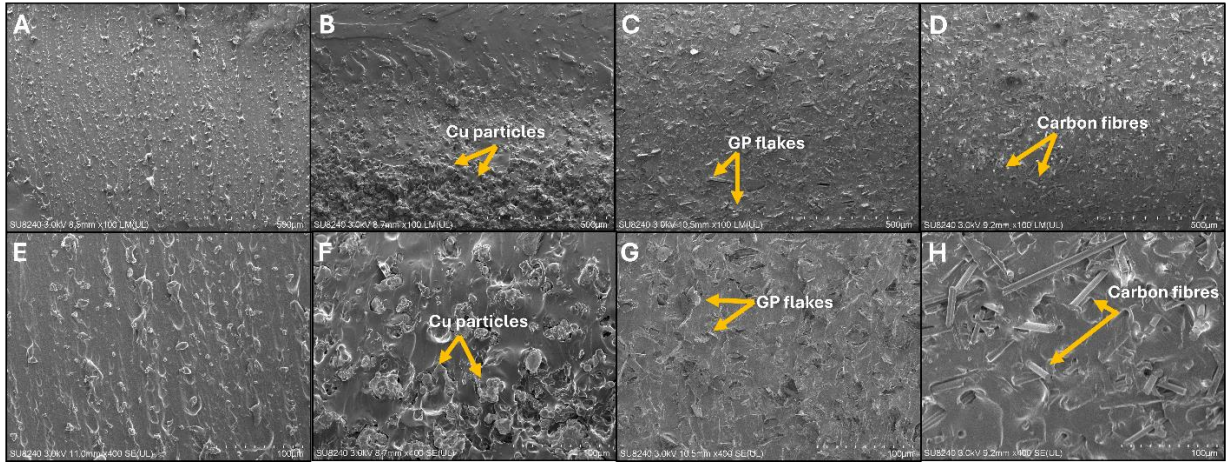

**Figure 3.** SEM images of loaded capacitors in this study with A-D) 100x magnification and E-H) 400x magnification. A and E) Pristine PDMS, B and F) PDMS-Cu, C and G) PDMS-GP, and D and H) PDMS-CF.

To evaluate the scalability of PDMS-CF composite capacitors, we fabricated parallel plate capacitors with areas of 2 cm², 4 cm² and 6 cm² and evaluated their area-normalized capacitance up to 200 MHz (Figure 4). The absolute capacitance values across the entire frequency range (Figure 4A) increased linearly with increasing area, indicating a uniform permittivity laterally through the composite. The average area-normalized capacitance is presented in Figure 4C and demonstrates that the capacitance of PDMS-CF composites are in the range of 10-14 pF up to 200 MHz. Additionally, the standard deviation in capacitance up to 200 MHz is under 6.2% of the average value, demonstrating that CF-loaded PDMS capacitors are well-mixed over a large area and behave as parallel plate capacitors as the size of the device is changed.

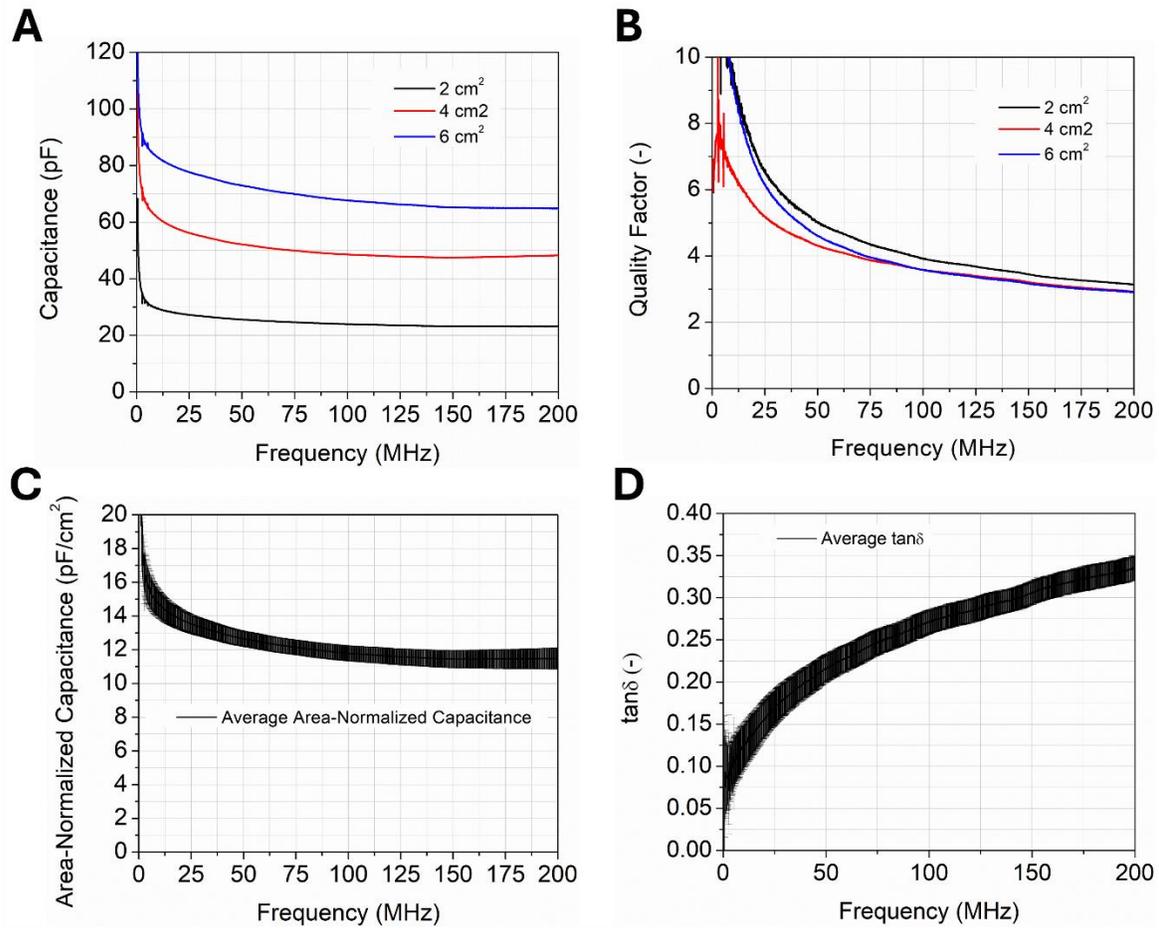

**Figure 4.** RF characterization of the sensing capacitors and their scalability: A) Capacitance and B) quality factor for PDMS-CF capacitors with area of 2 cm² – 6 cm² from 1 MHz to 200 MHz. C) area-normalized capacitance and D) area-normalized loss tangent (tanδ) of PDMS-CF capacitors with area of 2 cm² – 6 cm². Error bars in Figures C and D are for *n* = 3 area-normalized samples.

### 3.2 Broadband Dielectric Sensor Performance

The devices were characterized for their temperature sensitivity by placing them on a heating mantle and heating them in increments of 10°C in the range of 20°C – 110°C. In all cases, capacitance decreases as temperature increases. The plot of measured capacitance versus temperature in the range of 0.5. MHz to 200 MHz is shown for all four composites studied in this work in Figure 5. The sensitivity of the four composites as temperature sensors is reported in Table S1 of the Supporting Information. The inverse relationship between capacitance and temperature for PDMS and PDMS composites is attributed to a decrease in permittivity,[50] and thermal expansion of the PDMS.[51]

The capacitance of pristine PDMS devices in this work (Figure 5A) changes linearly with temperature with an average sensitivity of 0.18%·°C$^{-1}$ at all three investigated frequencies. PDMS-CF (Figure 5B) and PDMS-GP show slightly different kinetics in response to temperature, with the response of PDMS-CF capacitors saturating rapidly above 60°C at a relative response of over 80% and PDMS-GP following an exponential decay type response and starting to saturate above 90°C. The temperature response of PDMS-GP follows a similar trend to reported capacitors in the literature whose DC response followed exponential decay in the range of 30°C – 110°C for 15-25% graphite-loaded composite dielectrics.[22] For both PDMS-CF and PDMS-GP composites, we attribute the rapid initial relative response in capacitance to the thermal expansion of PDMS breaking of conductive pathways in the composite,[52] resulting in a rapid decrease in capacitance approaching that of PDMS. Finally, PDMS-Cu capacitors also demonstrated a nonlinear response at the three investigated frequencies, which indicates that some while some oxidation of the copper particles likely occurred after manufacturing due to the exposure of particle surfaces to air, some conductive non-oxidized copper is present in the composite.

Remarkably, the loaded composites did demonstrate frequency-dependant response to temperature, with operating PDMS-CF at 200 MHz resulting in an enhancement of 25% at 50°C and 15% above 60°C. Additionally, PDMS-GP consistently demonstrated a 3-5% enhancement in relative response operating at 10 MHz compared to 100 MHz. Unlike previous studies on PDMS-based capacitors operating at DC, this work demonstrates that an optimum operating frequency exists for capacitive temperature sensors, which has been explored in capacitive humidity sensors in the kHz range,[53] but has not previously been reported for PDMS composite-based temperature sensing materials. This temperature sensing approach demonstrates the limitations of DC sensing and provides further motivation for exploring the frequency-dependence of sensing materials.

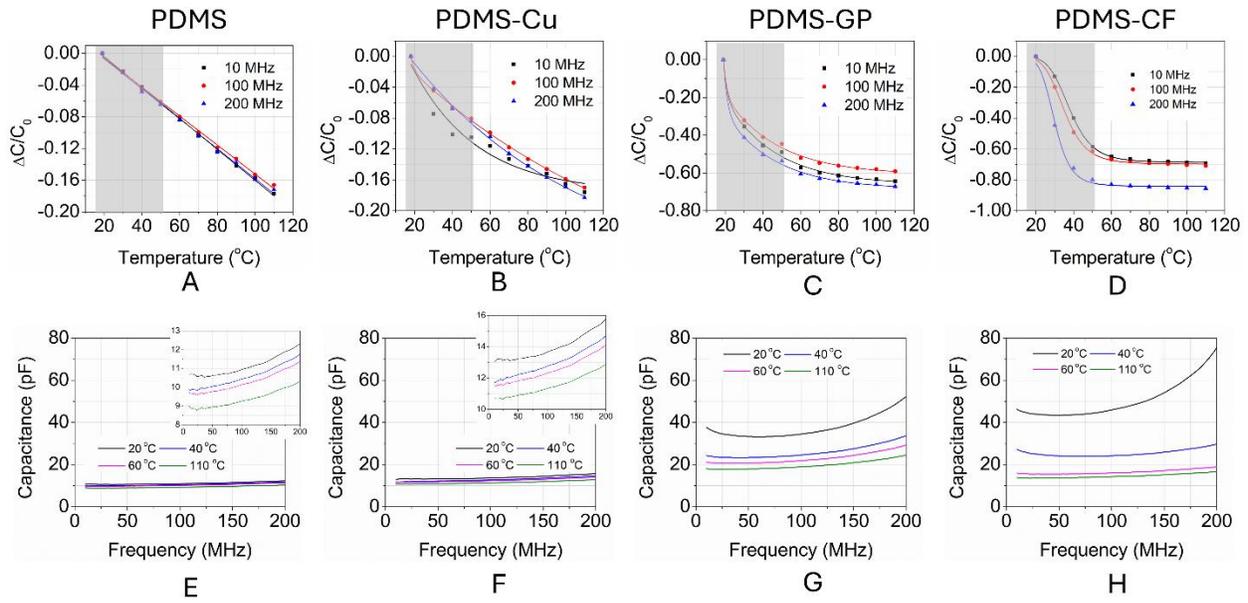

**Figure 5.** Frequency-dependant relative response to temperature of loaded composite capacitors A) pristine PDMS B) PDMS-CF C) PDMS-GP, and D) PDMS-Cu at different frequencies. The reference temperature for these measurements was taken at room temperature and the shaded region is the target sensing range.

### 3.3 Wireless and Chipless Temperature Sensing

To demonstrate the applicability of composite temperature sensors as passive measurement devices, we fabricated a sensing system based on coupled LC coils and took sensing measurements in the temperature range of 20°C – 110°C. To allow the temperature to be read wirelessly, the top coil ("Reader") was tuned to 6.78 MHz, the standard international, scientific, and medical (ISM) band and is an attractive frequency of operation from a regulatory-compliance and standardization perspective, widely used in inductive power transfer applications.[48,54] Pristine PDMS capacitors and PDMS-CF capacitors were sized and connected to a separate LC coil ("Sensor") to be tuned to the 6.78 MHz band. The reader and sensor coils were spaced by 45 mm, which achieves critical inductive coupling without observing multiple resonances (frequency splitting), which is common at small separation distances, due to over-coupling.[55] The temperature sensing setup is illustrated in **Figure 6A** with the simplified equivalent circuit diagram shown in Figure 6B. Changes in the reflection coefficient of the reader coils for wireless PDMS and PDMS-CF systems are shown in Figure 6C-D while changes in the reflection coefficient of sensor coils are shown in Figure 6F-G, respectively. In the wireless measurement of each capacitor, changes in the resonant frequencies of the sensor and reader coils are observed and quantified in Figure 6E and Figure 6H.

Observable at both the sensor and reader coils, the PDMS-CF-based sensor demonstrates a significant positive shift in the resonant frequency, which is attributed to the previously observed decrease in capacitance of the sensing material.[56] The change in the resonant frequency of the sensor coil results in the coils being poorly matched at 6.78 MHz and changes in both the amplitude, at resonance. Furthermore, an additional resonance is created by the changing capacitance detuning the sensor coil, creating a readable sensor response through $\Delta f_r$. This change in the resonant frequency can be quantified as the temperature response of the wireless system and has similarly been applied to measure various measurands including ice accumulation[32] and strain sensing.[33,57]

In the reader coil and for the PDMS-CF sensor, the resonant frequency increased by 36%, from 6.88 MHz to 9.41 MHz, compared to 0.9% for the pristine PDMS capacitor showing an increase from 6.63 MHz to 6.69 MHz. In the sensor coil itself, where the capacitor uses the PDMS-CF dielectric, the resonant frequency increased by 33.8% from 7.03 MHz to 9.41 MHz compared to 4.10% for PDMS whose resonant frequency increased from 7.50 MHz to 7.81 MHz.

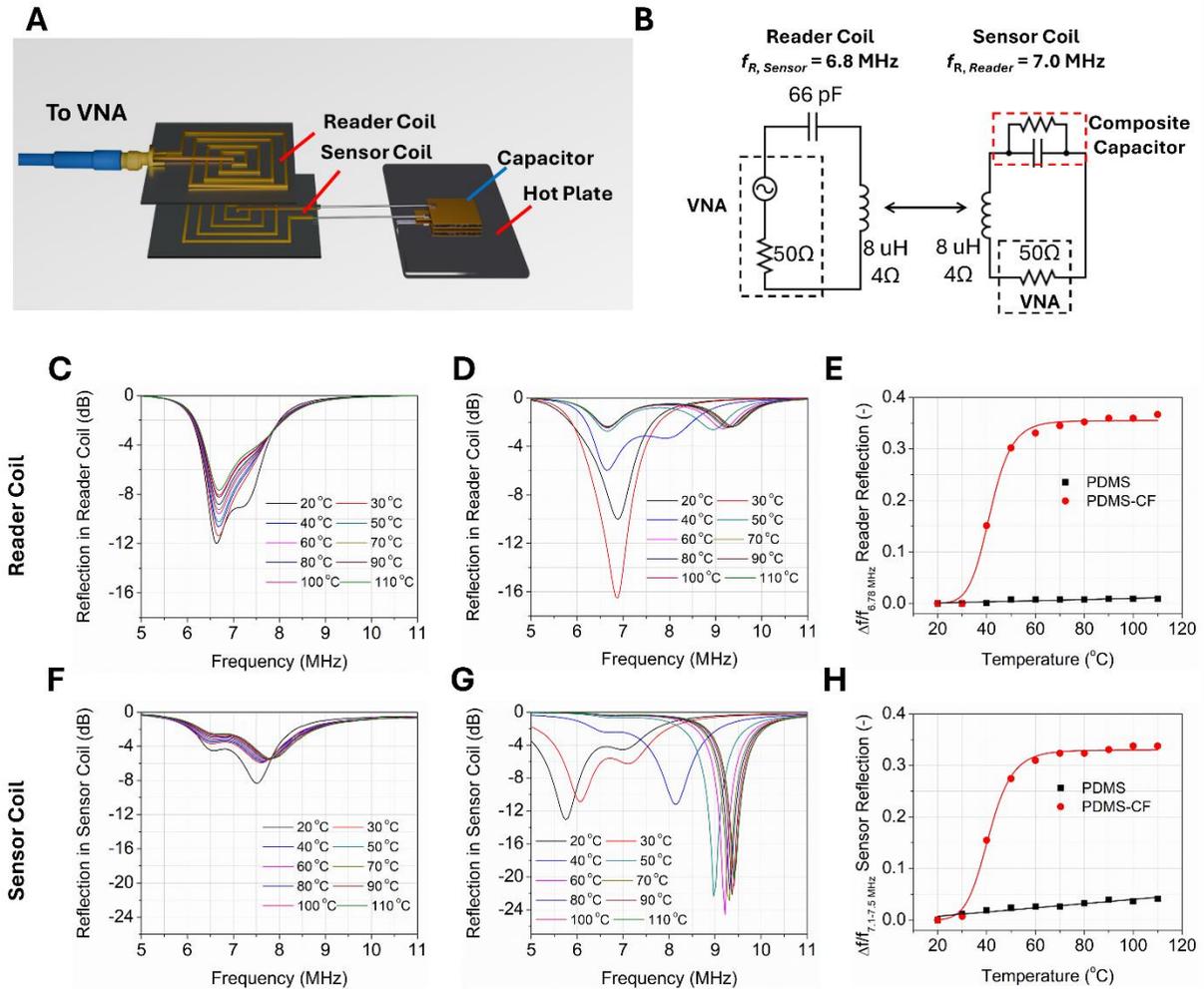

**Figure 6.** Chipless wireless readout using a tuned reader coil in the same band: A) 3D render of wireless temperature sensing experiment. B) Equivalent circuit of coupled RLC coils. C-H) temperature sensing data for wireless RLC coil system. Reflection coefficient data for reader coil ($S_{11}$) and sensor coil ($S_{22}$) for C and F) PDMS capacitor with an area of 16 cm² and D and G) PDMS-CF capacitor with an area of 4 cm². Relative response in the resonant frequency of the reflection in the E) reader coil and H) sensor coil.

While the frequency response of PDMS-CF wireless sensors significantly outperformed pristine PDMS wireless sensors, the resonant frequency change was negligible between 20°C and 30°C and only an increase in the amplitude of the reflection was observed, increasing from -10.0 dB to -16.5 dB. This is attributed to the very small change in permittivity, predominantly influencing $f_r$, compared to the change in the conductivity, which influences the equivalent series resistance and hence the losses in the capacitor. As such, the increase in reflection can be explained by occurred due to improved matching between the coils at 6.88 MHz as temperature increased from 20°C to 30°C, due to the improved matching between the source and 50 Ω termination on the sensing LC resonator. As the temperature rises further, the sensor coil is detuned, resulting in higher reflections at the unloaded resonant frequency.

This amplitude increase without resonant frequency yields a temperature sensor that is only reliable above 30°C. To address this, we adopted a novel readout mechanism where the sensor and reader coil are deliberately off-tuned, to isolate these two opposing effects. This was achieved by fabricating an identical wireless sensor system with a PDMS-CF capacitor having an area of 2 cm$^2$ (~23 pF), which resulted in a shift of the resonant frequency of the sensor coil from 7.03 MHz to 8.45 MHz.

The reflection of the reader coil of the system is shown in Figure 7A and the reflection of the sensor coil is shown in Figure 7B. The shift in resonant frequency of the sensor coil is measurable by the reader coil, which shows a distinct second resonance peak at 8.26 MHz. By changing the area of the PDMS-CF capacitor, the response in resonant frequency of both the sensor and reader coils can be read in the full temperature range of 20°C – 110°C and the relative response of the reader coil is compared to the pristine PDMS system in Figure 7D. Similar to the wireless system with the integrated 4 cm$^2$ PDMS-CF capacitor, the relative response of the 8.26 MHz resonance of the reader coil shows a positive shift with temperature to 10.96 MHz, which is a 32.7% change, and still significantly outperforms the wireless PDMS sensing system by over thirty six-fold. Additionally, deliberately de-tuning the LC resonators by adjusting the size of the capacitor enabled an increase in the sensing range of the reader coil while only sacrificing a negligible magnitude of the relative response read by the better matched system with a 4cm$^2$ PDMS-CF capacitor.

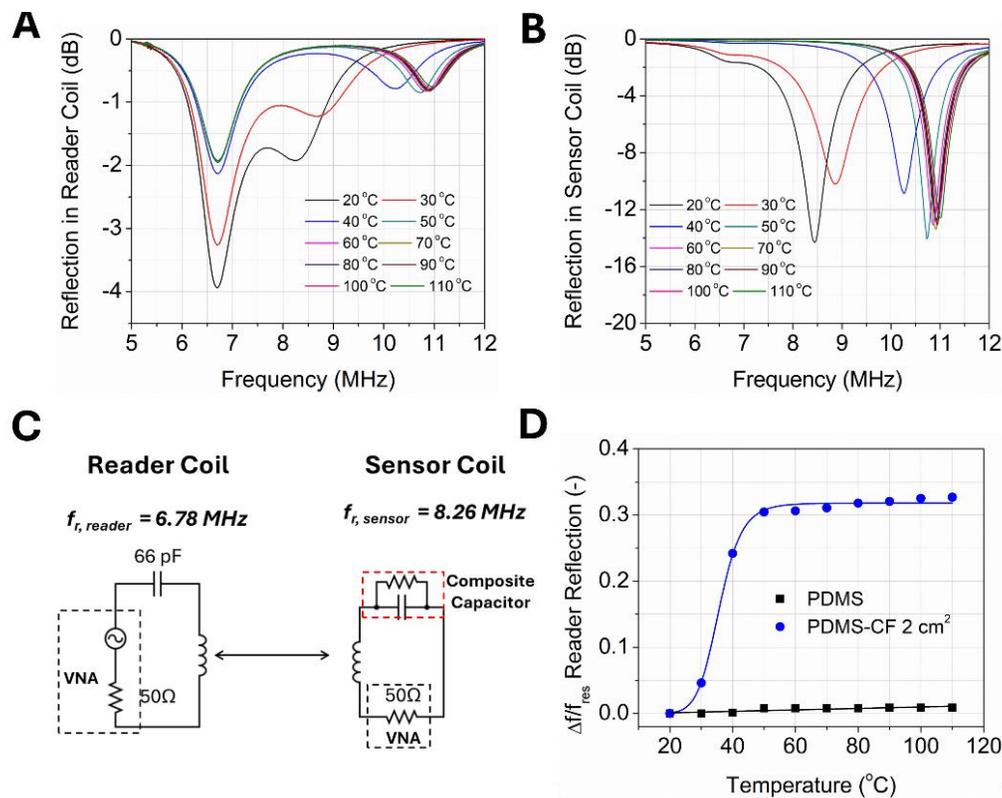

*Figure 7.* Chipless readout using an off-tuned reader coil: A) Reflection in reader coil and B) reflection in sensor coil for CF-PDMS capacitor with an area of 2 cm² in the range of 20°C – 110°C. C) Schematic diagram showing unloaded resonant frequency of reader coil and sensing coil with 2 cm² PDMS-CF capacitor. D) Relative response in resonant frequency of reflection in the reader coil with a reference 8.26 MHz for PDMS-CF compared to the resonant frequency ($f_R$) of PDMS at 6.63 MHz.

### *Wireless Sensor Readout Using a Portable Low-Cost NanoVNA*

To validate the wireless temperature sensing capability of the PDMS-CF composites under more practical test conditions, the sensor coil was measured using a lightweight, low-cost, and portable VNA, the open source NanoVNA,[46] (costing under $100, and occupying a smaller footprint than a phone) which can measure reflection coefficients in the range of 50 kHz – 1,500 MHz. The temperature sensing setup with the NanoVNA is illustrated in Figure 8**A** and the reflection coefficient at 30°C, 50°C and 110°C is shown in Figure 8B with the full temperature-dependant reflection and transmission coefficient data sets for the NanoVNA in Figure S8A and Figure S9 of the Supporting Information. The temperature sensing readout by the calibrated bench VNA and NanoVNA are visually similar across the full temperature range (Figure S8B), with both sets of traces showing the evolution of a second peak beyond 8 MHz as temperature increases. However, the NanoVNA readout shows a greater shift in the reflection coefficient than the PicoVNA readout at each temperature, which results in an increase in the observed relative response in temperature Figure 8C. The broadband reflection coefficient response is similar between the two measurement setups, suggesting

a good agreement between the portable and low-cost readout, and the standard VNA used in most studies.[29–36] The variations observed in the absolute change in the resonant frequency can be attributed to changes in the assembly of the measurement setup, where the separation and alignment between the inductively coupled coils.

These results demonstrate that, owing to its high sensitivity ($\Delta C, \Delta \tan \delta$) the PDMS-CF capacitive sensors can be integrated into portable, low-cost and chipless wireless temperature detection systems. The achieved sensitivity, compared with wireless temperature sensors from the literature in Figure 8D and E, and Table S2 (Supporting Information), demonstrates how the combination of engineered composites and sensitivity-driven readout frequency selection could achieve the most sensitive temperature detectors.

The average sensitivity of 0.55 %·°C$^{-1}$ over the range of 20°C – 110°C for the handheld VNA system outperforms state of the art radio frequency temperature sensors, including resonator-based RF temperature sensors based on CF-loaded capacitors with loadings of up to 40% CF by weight.[28] These previous sensors were reported based on engineered composites with comparable sensitivity (over 20% lower), the readout was only demonstrated at GHz frequencies, requiring more complex readout electronics, with no report of a chipless readout. Thus, the achieved sensitivity coupled with the demonstration of a chipless readout using a low-cost/open-source circuit create a pathway to widespread adoption of chipless frequency-domain sensors outside of a laboratory or without access to expensive specialty equipment.

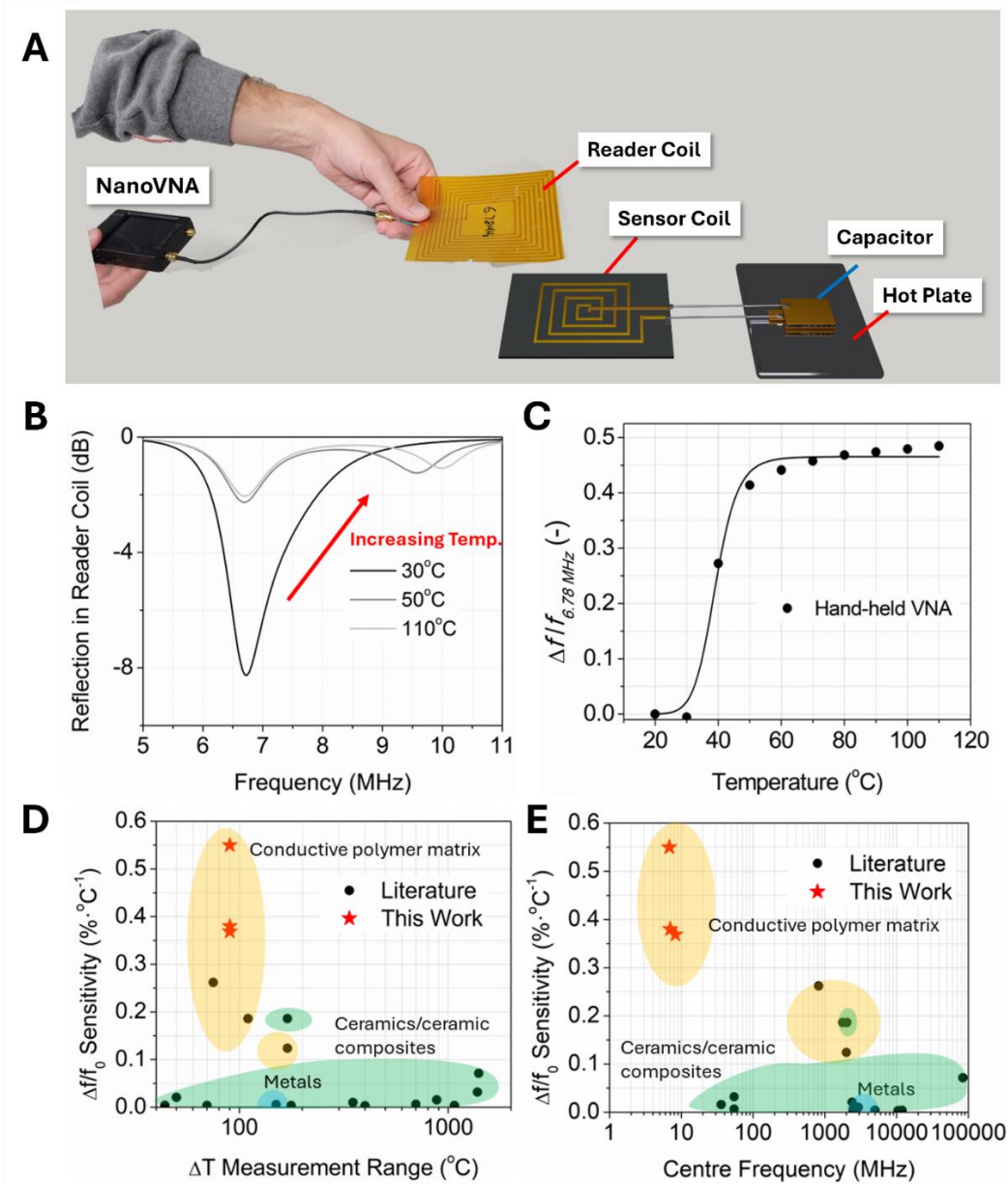

*Figure 8A. Chipless readout using inexpensive "handheld" VNA. A) Wireless sensing setup with NanoVNA connected to inductive coils and PDMS-CF parallel plate capacitor. B) Change in reader coil reflection as a function of temperature for PDMS-CF capacitor measured with the PicoVNA (solid lines) and NanoVNA (dashed lines). C) Relative response in resonant frequency of the reader coil as a function of temperature measured with low-cost handheld NanoVNA. D and E) Comparison with state-of-the-art frequency-domain temperature sensors in temperature measurement range and reference frequency (see Table S2 in Supporting Information for the data and references).*

## 4. Conclusion

We presented the first holistic investigation combining composite formulation, readout frequency optimisation, and a fully chipless demonstration of electromagnetic frequency-based temperature sensing. The influence of a variety of inexpensive and commercially available carbon and metallic powders as filler materials in PDMS composites for capacitive-based temperature sensors was explored. Loading PDMS with 10 vol% filler resulted in an increase in average room temperature capacitance from 3 pF for Cu powder (PDMS-Cu) to 25-35 pF for graphite powder (PDMS-GP) and milled carbon fiber powder (PDMS-CF). PDMS-GP and PDMS-CF demonstrated a remarkable enhancement in temperature sensitivity compared to pristine PDMS with a relative response in capacitance of up to 67.1% and 85.5%, respectively, at 200 MHz. Additionally, we demonstrated that the operating frequency of the capacitors is highly influential in the sensing response of the devices, resulting in a change in relative response of up to 15.5% in capacitance between 10 MHz and 200 MHz for PDMS-CF devices.

The overall system was demonstrated as a passive, wireless capacitor by integrating it into a system of coupled RLC coils tuned to 6.78 MHz, demonstrating its response in resonant frequency compared to a PDMS sensor tuned to the same room temperature resonant frequency. The resulting wireless PDMS-CF capacitor yielded a relative response in the resonant frequency of the reader coil of 33.8%, compared to 0.9% for PDMS, which is a 40x improvement in sensitivity. Using a novel frequency-shifted readout, a reliable temperature readout between 20°C and 110°C with a negligible change in relative response was achieved, improving the robustness beyond state-of-the-art passive LC sensors. The demonstrated sensitivity enabled the sensor to be interrogated using a low-cost and open-source VNA (the nanoVNA), in good agreement with calibrated instrumentation-grade lab equipment. These results demonstrate how large-area sensors based on scalable and low-cost processes could be sampled remotely without specialty equipment which could be adopted for widespread use.

## 5. Supporting Information

Schematic of PDMS composite capacitor temperature sensing setup. Area-normalized capacitance data for PDMS-CF. Sensitivity of composite capacitors at 10 MHz, 100 MHz and 200 MHz. Transmission data for wireless sensing experiments. NanoVNA temperature sensing setup and reflection data. RLC coil dimensions and inductance, resistance and quality factor calculations for fabricated coils.

## 6. Acknowledgements

The authors would like to acknowledge Dr. Joseph G. Manion (CGFigures) for their open access Blender asset library used in the production of Figures 1A, 6A and S1. This research made use of scikit-rf, an open-source Python package for RF and Microwave applications (https://www.scikit-rf.org/).[58]

**Funding:** This work was supported by the UK Engineering and Physical Sciences Research Council (EPSRC) through grants EP/W025752/1 and EP/Y002008/1, and by the UK Royal Society Research Grant RGS/R1/231028.

## 7. References

(1)     Lin, R.; Kim, H.; Achavananthadith, S.; Kurt, S. A.; Tan, S. C. C.; Yao, H.; Tee, B. C. K.; Lee, J. K. W.; Ho, J. S. Wireless Battery-Free Body Sensor Networks Using near-Field-Enabled Clothing. *Nat. Commun.* **2024**, *11*, 444. DOI: 10.1038/s41467-020-14311-2.

(2)     Liu, Z.; Tian, B.; Zhang, B.; Liu, J.; Zhang, Z.; Wang, S.; Luo, Y.; Zhao, L.; Shi, P.; Lin, Q.; Jiang, Z. A Thin- Film Temperature Sensor Based on a Flexible Electrode and Substrate. *Microsystems Nanoeng.* **2021**, *7*, 42. DOI: 10.1038/s41378-021-00271-0.

(3)     Wagih, M.; Balocchi, L.; Benassi, F.; Carvalho, N. B.; Chiao, J. C.; Correia, R.; Costanzo, A.; Cui, Y.; Georgiadou, D.; Gouveia, C.; Grosinger, J.; Ho, J. S.; Hu, K.; Komolafe, A.; Lemey, S.; Loss, C.; Marrocco, G.; Mitcheson, P.; Palazzi, V.; Panunzio, N.; Paolini, G.; Pinho, P.; Preishuber-Pflugl, J.; Qaragoez, Y.; Rahmani, H.; Rogier, H.; Lopera, J. R.; Roselli, L.; Schreurs, D.; Tentzeris, M.; Tian, X.; Torah, R.; Torres, R.; Van Torre, P.; Vital, D.; Beeby, S. Microwave-Enabled Wearables: Underpinning Technologies, Integration Platforms, and Next-Generation Roadmap. *IEEE J. Microwaves* **2023**, *3* (1), 193–226. DOI: 10.1109/JMW.2022.3223254.

(4)     Luo, Y.; Abidian, M. R.; Ahn, J. H.; Akinwande, D.; Andrews, A. M.; Antonietti, M.; Bao, Z.; Berggren, M.; Berkey, C. A.; Bettinger, C. J.; Chen, J.; Chen, P.; Cheng, W.; Cheng, X.; Choi, S. J.; Chortos, A.; Dagdeviren, C.; Dauskardt, R. H.; Di, C. A.; Dickey, M. D.; Duan, X.; Facchetti, A.; Fan, Z.; Fang, Y.; Feng, J.; Feng, X.; Gao, H.; Gao, W.; Gong, X.; Guo, C. F.; Guo, X.; Hartel, M. C.; He, Z.; Ho, J. S.; Hu, Y.; Huang, Q.; Huang, Y.; Huo, F.; Hussain, M. M.; Javey, A.; Jeong, U.; Jiang, C.; Jiang, X.; Kang, J.; Karnaushenko, D.; Khademhosseini, A.; Kim, D. H.; Kim, I. D.; Kireev, D.; Kong, L.; Lee, C.; Lee, N. E.; Lee, P. S.; Lee, T. W.; Li, F.; Li, J.; Liang, C.; Lim, C. T.; Lin, Y.; Lipomi, D. J.; Liu, J.; Liu, K.; Liu, N.; Liu, R.; Liu, Y.; Liu, Y.; Liu, Z.; Liu, Z.; Loh, X. J.; Lu, N.; Lv, Z.; Magdassi, S.; Malliaras, G. G.; Matsuhisa, N.; Nathan, A.; Niu, S.; Pan, J.; Pang, C.; Pei, Q.; Peng, H.; Qi, D.; Ren, H.; Rogers, J. A.; Rowe, A.; Schmidt, O. G.; Sekitani, T.; Seo, D. G.; Shen, G.; Sheng, X.; Shi, Q.; Someya, T.; Song, Y.; Stavrinidou, E.; Su, M.; Sun, X.; Takei, K.; Tao, X. M.; Tee, B. C. K.; Thean, A. V. Y.; Trung, T. Q.; Wan, C.; Wang, H.; Wang, J.; Wang, M.; Wang, S.; Wang, T.; Wang, Z. L.; Weiss, P. S.; Wen, H.; Xu, S.; Xu, T.; Yan, H.; Yan, X.; Yang, H.; Yang, L.; Yang, S.; Yin, L.; Yu, C.; Yu, G.; Yu, J.; Yu, S. H.; Yu, X.; Zamburg, E.; Zhang, H.; Zhang, X.; Zhang, X.; Zhang, X.; Zhang, Y.; Zhang, Y.; Zhao, S.; Zhao, X.; Zheng, Y.; Zheng, Y. Q.; Zheng, Z.; Zhou, T.; Zhu, B.; Zhu, M.; Zhu, R.; Zhu, Y.; Zhu, Y.; Zou, G.; Chen, X. Technology Roadmap for Flexible Sensors. *ACS Nano* **2023**, *17* (6), 5211–5295. DOI: 10.1021/acsnano.2c12606.

(5)     Liu, H.; Li, Q.; Zhang, S.; Yin, R.; Liu, X.; He, Y. Electrically Conductive Polymer Composites for Smart Flexible Strain Sensors : A Critical Review. *J. Mater. Chem. C*

**2018**, *6* (45), 12121–12141. DOI: 10.1039/c8tc04079f.

(6)     Dai, K.; Liu, C.; Guo, Z. Lightweight Conductive Graphene/Thermoplastic Polyurethane Foams with Ultrahigh Compressibility for Piezoresistive Sensing. *J. Mater. Chem. C* **2017**, *5* (1), 73–83. DOI: 10.1039/c6tc03713e.

(7)     Lee, J.; Lim, M.; Yoon, J.; Kim, M. S.; Choi, B.; Kim, D. M.; Kim, D. H.; Park, I.; Choi, S. Transparent, Flexible Strain Sensor Based on a Solution-Processed Carbon Nanotube Network. *ACS Appl. Mater. Interfaces* **2017**, *9* (31), 26279–26285. DOI: 10.1021/acsami.7b03184.

(8)     Liu, H.; Huang, W.; Gao, J.; Dai, K.; Zheng, G.; Liu, C.; Shen, C.; Yan, X.; Guo, J.; Guo, Z. Piezoresistive Behavior of Porous Carbon Nanotube-Thermoplastic Polyurethane Conductive Nanocomposites with Ultrahigh Compressibility. *Appl. Phys. Lett.* **2016**, *108* (1), 011904. DOI: 10.1063/1.4939265.

(9)     Ambrosetti, G.; Grimaldi, C.; Balberg, I.; Maeder, T.; Danani, A.; Ryser, P. Solution of the Tunneling-Percolation Problem in the Nanocomposite Regime. *Phys. Rev. B* **2010**, *81* (15), 155434. DOI: 10.1103/PhysRevB.81.155434.

(10)    Chuang, W.-J.; Chiu, W.-Y.; Tai, H.-J. Temperature-Dependent Conductive Composites: Poly(N-Isopropylacrylamide-Co-N-Methylol Acrylamide) and Carbon Black Composite Films. *J. Mater. Chem.* **2012**, *22* (38), 20311. DOI: 10.1039/c2jm33601d.

(11)    Chen, Z.; Zhao, D.; Rui, M.; Zhang, X.; Rao, J.; Yin, Y.; Wang, X.; Yi, F. Flexible Temperature Sensors Based on Carbon Nanomaterials. *J. Mater. Chem. B* **2021**, *9* (8), 1941–1964. DOI: 10.1039/d0tb02451a.

(12)    Zhang, H.; Chen, X.; Liu, Y.; Yang, C.; Liu, W.; Qi, M.; Zhang, D. PDMS Film-Based Flexible Pressure Sensor Array with Surface Protruding Structure for Human Motion Detection and Wrist Posture Recognition. *ACS Appl. Mater. Interfaces* **2024**, *16* (2), 2554–2563. DOI: 10.1021/acsami.3c14036.

(13)    Masihi, S.; Panahi, M.; Maddipatla, D.; Hanson, A. J.; Bose, A. K.; Hajian, S.; Palaniappan, V.; Narakathu, B. B.; Bazuin, B. J.; Atashbar, M. Z. Highly Sensitive Porous PDMS-Based Capacitive Pressure Sensors Fabricated on Fabric Platform for Wearable Applications. *ACS Sensors* **2021**, *6* (3), 938–949. DOI: 10.1021/acssensors.0c02122.

(14)    Larmagnac, A.; Eggenberger, S.; Janossy, H.; Vörös, J. Stretchable Electronics Based on Ag-PDMS Composites. *Sci. Rep.* **2014**, *4*, 7294. DOI: 10.1038/srep07254.

(15)    Luo, R.; Li, H.; Du, B.; Zhou, S.; Zhu, Y. A Simple Strategy for High Stretchable, Flexible and Conductive Polymer Films Based on PEDOT:PSS-PDMS Blends. *Org. Electron.* **2020**, *76*, 105451. DOI: 10.1016/j.orgel.2019.105451.

(16)    Hwang, J.; Kim, Y.; Yang, H.; Oh, J. H. Fabrication of Hierarchically Porous Structured PDMS Composites and Their Application as a Flexible Capacitive Pressure Sensor.

*Compos. Part B Eng.* **2021**, *211*, 108607. DOI: 10.1016/j.compositesb.2021.108607.

(17) Zeng, P.; Pan, P.; He, J.; Yang, Z.; Song, H.; Zhang, J. Porous Composite PDMS for a Pressure Sensor with a Wide Linear Range. *ACS Appl. Nano Mater.* **2024**, *7* (1), 455–465. DOI: 10.1021/acsanm.3c04610.

(18) Guo, X.; Liu, T.; Tang, Y.; Li, W.; Liu, L.; Wang, D.; Zhang, Y.; Zhang, T.; Zhu, X.; Guan, Y.; Li, X.; Chen, Y.; Wu, X.; Xiao, G.; Wang, X.; Zhang, R.; Wang, D.; Mai, Z.; Hong, W.; Hong, Q.; Zhao, Y.; Zhang, Y.; Wang, M.; Yan, F.; Xing, G. Bioinspired Low Hysteresis Flexible Pressure Sensor Using Nanocomposites of Multiwalled Carbon Nanotubes, Silicone Rubber, and Carbon Nanofiber for Human-Computer Interaction. *ACS Appl. Nano Mater.* **2024**, *7* (13), 15626–15639. DOI: 10.1021/acsanm.4c02631.

(19) Zhu, W. Bin; Wang, Y. Y.; Fan, T.; Zhu, Y.; Tang, Z. H.; Huang, P.; Li, Y. Q.; Fu, S. Y. Comprehensive Investigation of the Temperature-Dependent Electromechanical Behaviors of Carbon Nanotube/Polymer Composites. *Langmuir* **2024**, *40* (15), 8170–8179. DOI: 10.1021/acs.langmuir.4c00231.

(20) Müller, A.; Wapler, M. C.; Wallrabe, U. A Quick and Accurate Method to Determine the Poisson's Ratio and the Coefficient of Thermal Expansion of PDMS. *Soft Matter* **2019**, *15* (4), 779–784. DOI: 10.1039/c8sm02105h.

(21) Chuang, H. S.; Wereley, S. Design, Fabrication and Characterization of a Conducting PDMS for Microheaters and Temperature Sensors. *J. Micromechanics Microengineering* **2009**, *19* (4), 045010. DOI: 10.1088/0960-1317/19/4/045010.

(22) Shih, W. P.; Tsao, L. C.; Lee, C. W.; Cheng, M. Y.; Chang, C.; Yang, Y. J.; Fan, K. C. Flexible Temperature Sensor Array Based on a Graphite-Polydimethylsiloxane Composite. *Sensors* **2010**, *10* (4), 3597–3610. DOI: 10.3390/s100403597.

(23) Oppili Prasad, L.; Sreelal Pillai, S.; Sambandan, S. Micro-Strain and Temperature Sensors for Space Applications with Graphite-PDMS Composite. *FLEPS 2019 - IEEE Int. Conf. Flex. Printable Sensors Syst. Proc.* **2019**, 13–15. DOI: 10.1109/FLEPS.2019.8792312.

(24) Luo, J.; Liu, G. S.; Zhou, W.; Hu, S.; Chen, L.; Chen, Y.; Luo, Y.; Chen, Z. A Graphene-PDMS Hybrid Overcoating Enhanced Fiber Plasmonic Temperature Sensor with High Sensitivity and Fast Response. *J. Mater. Chem. C* **2020**, *8* (37), 12893–12901. DOI: 10.1039/d0tc02350g.

(25) Wu, L.; Qian, J.; Peng, J.; Wang, K.; Liu, Z.; Ma, T.; Zhou, Y.; Wang, G.; Ye, S. Screen-Printed Flexible Temperature Sensor Based on FG/CNT/PDMS Composite with Constant TCR. *J. Mater. Sci. Mater. Electron.* **2019**, *30* (10), 9593–9601. DOI: 10.1007/s10854-019-01293-1.

(26) Liu, H.; Sun, K.; Guo, X. L.; Liu, Z. L.; Wang, Y. H.; Yang, Y.; Yu, D.; Li, Y. T.; Ren, T. L. An Ultrahigh Linear Sensitive Temperature Sensor Based on PANI:Graphene and PDMS Hybrid with Negative Temperature Compensation. *ACS Nano* **2022**, *16* (12),

21527–21535. DOI: 10.1021/acsnano.2c10342.

(27) Potyrailo, R. A.; Go, S.; Sexton, D.; Li, X.; Alkadi, N.; Kolmakov, A.; Amm, B.; St-Pierre, R.; Scherer, B.; Nayeri, M.; Wu, G.; Collazo-Davila, C.; Forman, D.; Calvert, C.; Mack, C.; McConnell, P. Extraordinary Performance of Semiconducting Metal Oxide Gas Sensors Using Dielectric Excitation. *Nat. Electron.* **2020**, *3* (5), 280–289. DOI: 10.1038/s41928-020-0402-3.

(28) Wagih, M.; Shi, J.; Li, M.; Komolafe, A.; Whittaker, T.; Schneider, J.; Kumar, S.; Whittow, W.; Beeby, S. Wide-Range Soft Anisotropic Thermistor with a Direct Wireless Radio Frequency Interface. *Nat. Commun.* **2024**, *15*, 452. DOI: 10.1038/s41467-024-44735-z.

(29) Wagih, M.; Shi, J. Toward the Optimal Antenna-Based Wireless Sensing Strategy: An Ice Sensing Case Study. *IEEE Open J. Antennas Propag.* **2022**, *3*, 687–699. DOI: 10.1109/OJAP.2022.3182770.

(30) Kazemi, K. K.; Zarifi, T.; Mohseni, M.; Narang, R.; Golovin, K.; Zarifi, M. H. Smart Superhydrophobic Textiles Utilizing a Long-Range Antenna Sensor for Hazardous Aqueous Droplet Detection plus Prevention. *ACS Appl. Mater. Interfaces* **2021**, *13* (29), 34877–34888. DOI: 10.1021/acsami.1c07880.

(31) Wiltshire, B. D.; Alijani, M.; Sopha, H.; Pavliňák, D.; Hromadko, L.; Zazpe, R.; Thalluri, S. M.; Kolibalova, E.; Macak, J. M.; Zarifi, M. H. Gigahertz-Based Visible Light Detection Enabled via CdS-Coated TiO2 Nanotube Layers. *ACS Appl. Mater. Interfaces* **2023**, *15* (14), 18379–18390. DOI: 10.1021/acsami.2c21877.

(32) Kozak, R.; Wiltshire, B. D.; Khandoker, M. A. R.; Golovin, K.; Zarifi, M. H. Modified Microwave Sensor with a Patterned Ground Heater for Detection and Prevention of Ice Accumulation. *ACS Appl. Mater. Interfaces* **2020**, *12* (49), 55483–55492. DOI: 10.1021/acsami.0c17173.

(33) Dijvejin, Z. A.; Kazemi, K. K.; Alasvand Zarasvand, K.; Zarifi, M. H.; Golovin, K. Kirigami-Enabled Microwave Resonator Arrays for Wireless, Flexible, Passive Strain Sensing. *ACS Appl. Mater. Interfaces* **2020**, *12* (39), 44256–44264. DOI: 10.1021/acsami.0c10384.

(34) Singh, S. K.; Azad, P.; Akhtar, M. J.; Kar, K. K. Improved Methanol Detection Using Carbon Nanotube-Coated Carbon Fibers Integrated with a Split-Ring Resonator-Based Microwave Sensor. *ACS Appl. Nano Mater.* **2018**, *1* (9), 4746–4755. DOI: 10.1021/acsanm.8b00965.

(35) Marasco, I.; Niro, G.; Marzo, G. De; Rizzi, F.; D, A.; Grande, M.; Vittorio, M. De; Member, S. Design and Fabrication of a Plastic-Free Antenna on a Sustainable Chitosan Substrate. *IEEE Electron Device Lett.* **2023**, *44* (2), 341–344. DOI: 10.1109/LED.2022.3232986.

(36) Moradpour, M.; Jain, M. C.; Tanguy, N. R.; Colegrave, K.; Zarifi, M. H. Exploring

PEDOT : PSS Interaction with Hazardous Gas Molecules in Microwave Regime Using Organic Microwave Resonators. *Chem. Eng. J.* **2023**, *458*, 141500. DOI: 10.1016/j.cej.2023.141500.

(37) Andre, R. S.; Schneider, R.; DeLima, G. R.; Fugikawa-Santos, L.; Correa, D. S. Wireless Sensor for Meat Freshness Assessment Based on Radio Frequency Communication. *ACS Sensors* **2024**, *9* (2), 631–637. DOI: 10.1021/acssensors.3c01657.

(38) Carr, A. R.; Chan, Y. J.; Reuel, N. F. Contact-Free, Passive, Electromagnetic Resonant Sensors for Enclosed Biomedical Applications: A Perspective on Opportunities and Challenges. *ACS Sensors* **2023**, *8* (3), 943–955. DOI: 10.1021/acssensors.2c02552.

(39) Kalhori, A. H.; Kim, W. S. Printed Wireless Sensing Devices Using Radio Frequency Communication. *ACS Appl. Electron. Mater.* **2022**, 5 (1), 1-10. DOI: 10.1021/acsaelm.2c01374.

(40) Li, X.; Chen, W.; Li, H.; Shen, B.; He, J.; Gao, H.; Bin, F.; Li, H.; Xiao, D. Temperature Self-Compensating Intelligent Wireless Measuring Contact Lens for Quantitative Intraocular Pressure Monitoring. *ACS Appl. Mater. Interfaces* **2024**, *16* (17), 22522–22531. DOI: 10.1021/acsami.4c02289.

(41) Reuel, N. F.; McAuliffe, J. C.; Becht, G. A.; Mehdizadeh, M.; Munos, J. W.; Wang, R.; Delaney, W. J. Hydrolytic Enzymes as (Bio)-Logic for Wireless and Chipless Biosensors. *ACS Sensors* **2016**, *1* (4), 348–353. DOI: 10.1021/acssensors.5b00259.

(42) Dong, Z.; Li, Z.; Yang, F.; Qiu, C. W.; Ho, J. S. Sensitive Readout of Implantable Microsensors Using a Wireless System Locked to an Exceptional Point. *Nat. Electron.* **2019**, *2* (8), 335–342. DOI: 10.1038/s41928-019-0284-4.

(43) Yang, M.; Ye, Z.; Alsaab, N.; Farhat, M.; Chen, P. Y. In-Vitro Demonstration of Ultra-Reliable, Wireless and Batteryless Implanted Intracranial Sensors Operated on Loci of Exceptional Points. *IEEE Trans. Biomed. Circuits Syst.* **2022**, *16* (2), 287–295. DOI: 10.1109/TBCAS.2022.3164697.

(44) Wagih, M.; Komolafe, A.; Hillier, N. Screen-Printable Flexible Textile-Based Ultra-Broadband Millimeter-Wave DC-Blocking Transmission Lines Based on Microstrip-Embedded Printed Capacitors. *IEEE J. Microwaves* **2022**, *2* (1), 162–173. DOI: 10.1109/JMW.2021.3126927.

(45) Mariotti, C.; Cook, B. S.; Roselli, L.; Tentzeris, M. M. State-of-the-Art Inkjet-Printed Metal-Insulator-Metal (MIM) Capacitors on Silicon Substrate. *IEEE Microw. Wirel. Components Lett.* **2015**, *25* (1), 13–15. DOI: 10.1109/LMWC.2014.2365745.

(46) NanoVNA | Very tiny handheld Vector Network Analyzer https://nanovna.com/.

(47) Li, J.; Li, Y.; Wang, Z.; Bian, H.; Hou, Y.; Wang, F.; Xu, G.; Liu, B.; Liu, Y. Ultrahigh Oxidation Resistance and High Electrical Conductivity in Copper-Silver Powder. *Sci.*

*Rep.* **2016**, *6*, 39650. DOI: 10.1038/srep39650.

(48) Wagih, M.; Komolafe, A.; Ullah, I. A Wearable All-Printed Textile-Based 6 .78 MHz 15 W-Output Wireless Power Transfer System and Its Screen-Printed Joule Heater Application. *IEEE Trans. Ind. Electron.* **2024**, *71* (4), 3741–3750. DOI: 10.1109/TIE.2023.3277112.

(49) Xie, Z.; Liu, D.; Xiao, Y.; Wang, K.; Zhang, Q.; Wu, K.; Fu, Q. The Effect of Filler Permittivity on the Dielectric Properties of Polymer-Based Composites. *Compos. Sci. Technol.* **2022**, *222*, 109342. DOI: 10.1016/j.compscitech.2022.109342.

(50) Sharma, P. K.; Gupta, N.; Dankov, P. I. Analysis of Dielectric Properties of Polydimethylsiloxane (PDMS) as a Flexible Substrate for Sensors and Antenna Applications. *IEEE Sens. J.* **2021**, *21* (17), 19492–19504. DOI: 10.1109/JSEN.2021.3089827.

(51) Kim, B.; Park, M.; Kim, Y. S.; Jeong, U. Thermal Expansion and Contraction of an Elastomer Stamp Causes Position-Dependent Polymer Patterns in Capillary Force Lithography. *ACS Appl. Mater. Interfaces* **2011**, *3* (12), 4695–4702. DOI: 10.1021/am201118u.

(52) Yang, Z.; Peng, H.; Wang, W.; Liu, T. Temperature Dependence of the Conductivity Behavior of Graphite Nanoplatelet-Filled Epoxy Resin Composites. *J. Appl. Polym. Sci.* **2009**, *113*, 1515–1519. DOI: 10.1002/app.

(53) Islam, T.; Ur Rahman, M. Z. Investigation of the Electrical Characteristics on Measurement Frequency of a Thin-Film Ceramic Humidity Sensor. *IEEE Trans. Instrum. Meas.* **2016**, *65* (3), 694–702. DOI: 10.1109/TIM.2015.2506302.

(54) Ziegelberger, G.; Croft, R.; Feychting, M.; Green, A. C.; Hirata, A.; d'Inzeo, G.; Jokela, K.; Loughran, S.; Marino, C.; Miller, S.; Oftedal, G.; Okuno, T.; van Rongen, E.; Röösli, M.; Sienkiewicz, Z.; Tattersall, J.; Watanabe, S. *Guidelines for Limiting Exposure to Electromagnetic Fields (100 KHz to 300 GHz)*, *Health Physics* **2020**, 118 (5) 483-524. DOI: 10.1097/HP.0000000000001210.

(55) Imura, T.; Hori, Y. Maximizing Air Gap and Efficiency of Magnetic Resonant Coupling for Wireless Power Transfer Using Equivalent Circuit and Neumann Formula. *IEEE Trans. Ind. Electron.* **2011**, *58* (10), 4746–4752. DOI: 10.1109/TIE.2011.2112317.

(56) Ji, Y.; Tan, Q.; Wang, H.; Lv, W.; Dong, H.; Xiong, J. A Novel Surface LC Wireless Passive Temperature Sensor Applied in Ultra-High Temperature Measurement. *IEEE Sens. J.* **2019**, *19* (1), 102–112. DOI: 10.1109/JSEN.2018.2872915.

(57) Song, L.; Myers, A. C.; Adams, J. J.; Zhu, Y. Stretchable and Reversibly Deformable Radio Frequency Antennas Based on Silver Nanowires. *ACS Appl. Mater. Interfaces* **2014**, *6* (6), 4248–4253. DOI: 10.1021/am405972e.

(58) Arsenovic, A.; Hillairet, J.; Anderson, J.; Forsten, H.; Ries, V.; Eller, M.; Sauber, N.; Weikle, R.; Barnhart, W.; Forstmayr, F. Scikit-Rf: An Open Source Python Package for Microwave Network Creation, Analysis, and Calibration [Speaker's Corner]. *IEEE Microw. Mag.* **2022**, *23* (1), 98–105. DOI: 10.1109/MMM.2021.3117139.

# Large-Area Conductor-Loaded PDMS Dielectric Composites for High-Sensitivity Wireless and Chipless Electromagnetic Temperature Sensors

Benjamin King, Nikolas Bruce, and Mahmoud Wagih

University of Glasgow, James Watt School of Engineering, Glasgow, UK

## *Supporting Information*

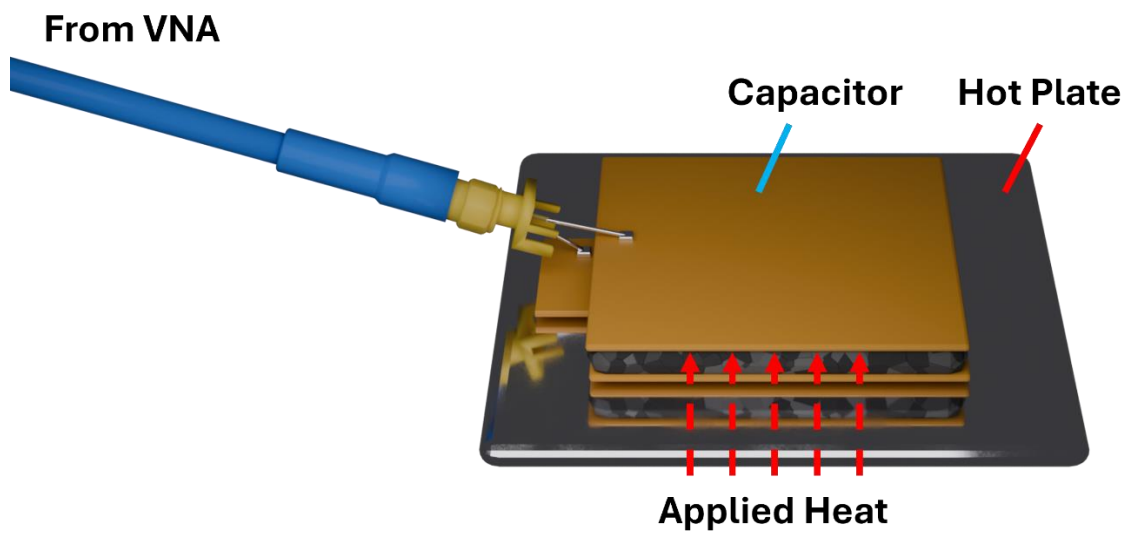

*Figure S1. Schematic of PDMS composite capacitor temperature sensing work bench.*

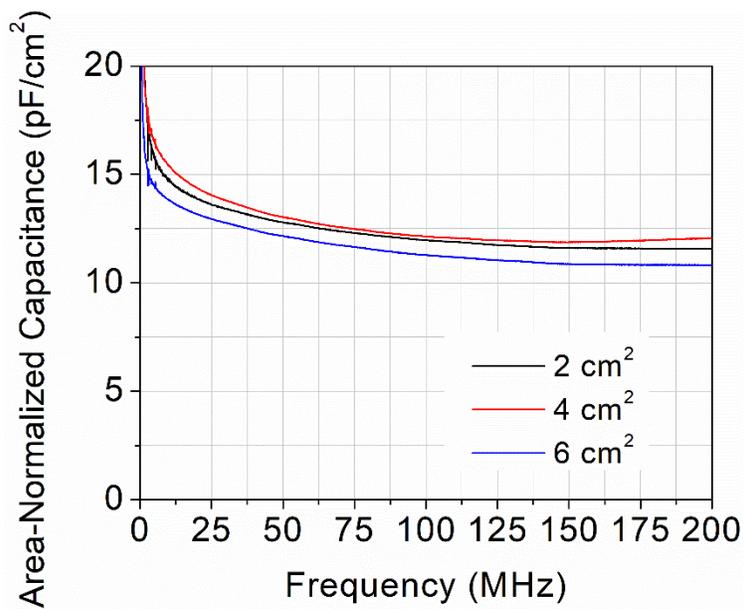

*Figure S2. Area-normalized capacitance for PDMS-CF capacitors of 2 cm² (black line), 4 cm² (red line) and 6 cm² (blue line) from 10 MHz to 200 MHz.*

**Table S1.** Mean sensing response and sensitivity for loaded composite capacitors.

| Sample ID | Filler | $S$ x10$^{-2}$ at 10 MHz (MHz·°C$^{-1}$) | $S$ x10$^{-2}$ at 100 MHz (MHz·°C$^{-1}$) | $S$ x10$^{-2}$ at 200 MHz (MHz·°C$^{-1}$) |
|---|---|---|---|---|
| Pristine | None | 0.14 | 0.16 | 0.18 |
| PDMS-CF* | Milled CF | 1.99 | 2.17 | 2.35 |
| PDMS-GP* | Graphite Powder | 1.31 | 1.08 | 1.39 |
| PDMS-Cu | Cu Metal Powder | 0.16 | 0.18 | 0.19 |

*Sensitivity data taken from 20 °C to 50 °C.

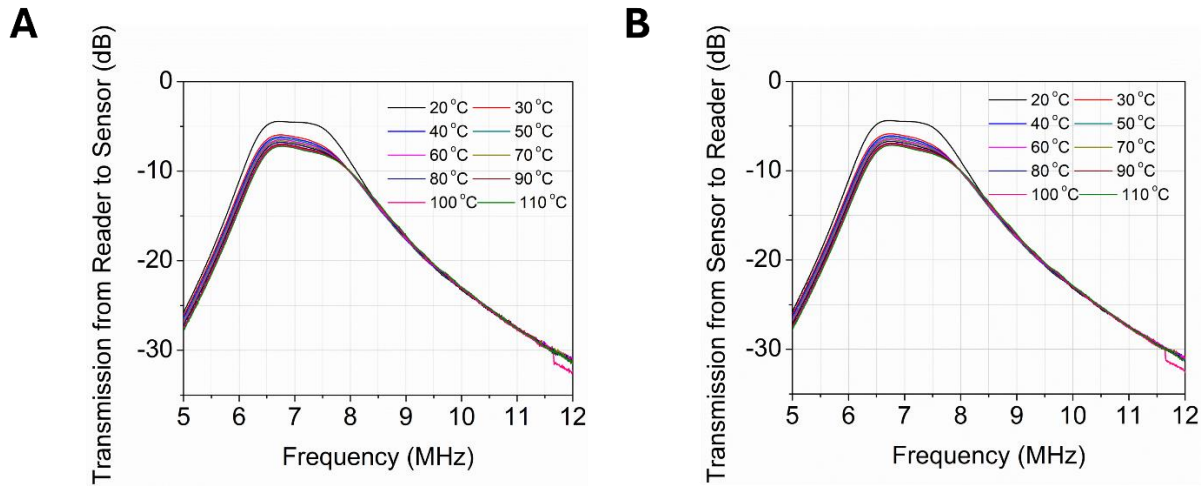

*Figure S3. The PDMS sensor's S21 response, indicating minimal amplitude or frequency variations A) S12 transmission coefficient and B) S21 transmission coefficient data for 16 cm² active area wireless PDMS capacitor sensor with a distance between coils of 45 mm.*

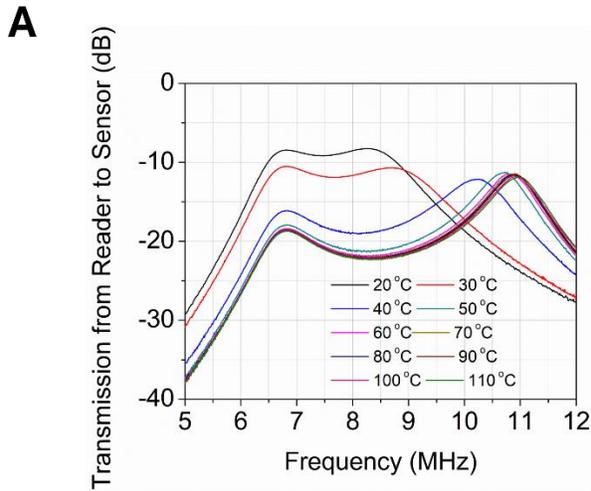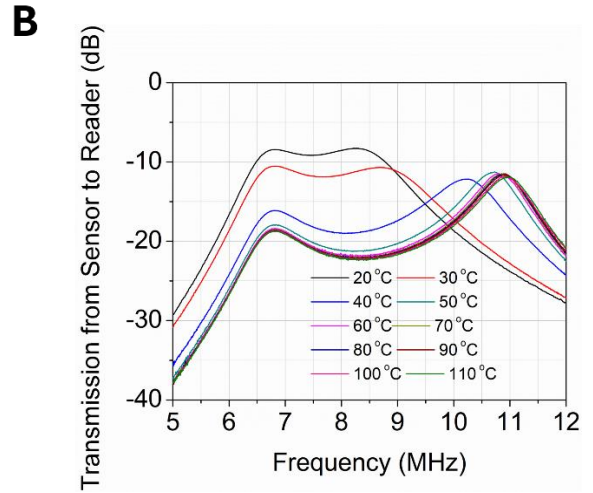

*Figure S4. The 2 cm² PDMS-CF sensor's transmission response A) S12 transmission coefficient and B) S21 transmission coefficient data for wireless 2cm² active area PDMS-CF capacitor sensor with a distance between coils of 45 mm.*

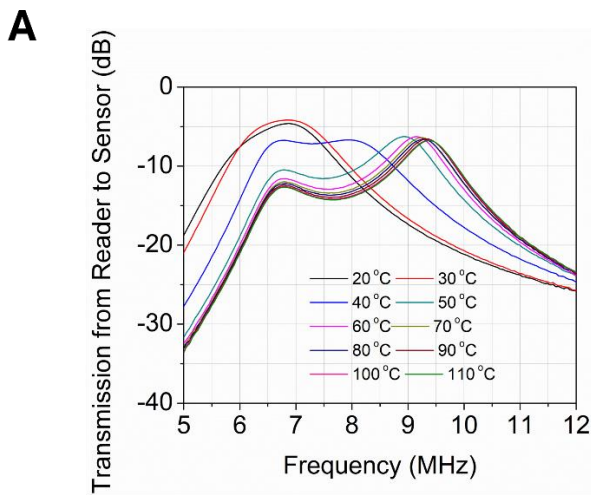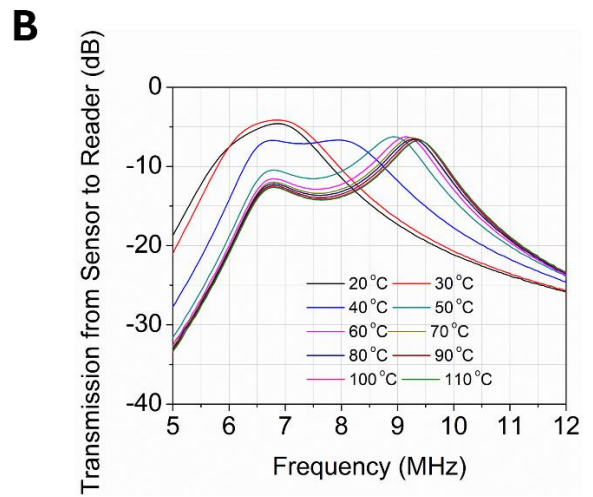

*Figure S5. The 4 cm² PDMS-CF sensor's transmission response A) S12 transmission coefficient and B) S21 transmission coefficient data for 4cm² active area PDMS-CF capacitor sensor with a distance between coils of 45 mm.*

*Table S2. Comparison with previously reported frequency-domain and AC/RF temperature sensors.*

| Reference | Sensor Architecture | Active Material | Temperature Sensing Range $\Delta T$ (°C) | Centre Frequency, $f_0$ (MHz) | Sesitivity (MHz·°C$^{-1}$) | Frequency-Normalized Sensitivity (%·°C$^{-1}$) |
|---|---|---|---|---|---|---|
| S1 | Dielectric resonator (chipless tag) | Ceramic | 20 – 370 | 2920 | 0.307 | 0.0102 |
| S2 | Chipless split-ring resonator | Ceramic | 28 - 1100 | 2470 | 0.0956 | 0.00395 |
| S3 | Near-field LC resonator | Ceramic | 19 – 900 | 36 | 0.00522 | 0.0158 |
| S4 | Dual split-ring resonator | Ceramic | 23 – 200 | 11930 | 0.462 | 0.00389 |
| S5 | Dual split-ring resonator | Ceramic | 25 – 135 | 2030 | 0.205 | 0.186 |
| S6 | Patch antenna | Polymer | 33 – 77 | 5000 | 0.205 | 0.0041 |
| S7 | Microstrip resonator | Rogers 3210 | 30 – 80 | 2400 | 0.5 | 0.0208 |
| S8 | Bulk acoustic wave resonator | Ceramic | 10 – 80 | 2480 | 0.1 | 0.00403 |
| S9 | LC Resonator | Ceramic | 500 – 1200 | 54.5 | 0.00357 | 0.00655 |
| S10 | Cylindrical antenna / resonator | Ceramic | 0 – 400 | 10500 | 0.35 | 0.00333 |
| S11 | RF cavity | Metal | 20 – 170 | 2493 | 0.005 | 0.006 |
| S12 | Patch antenna | Conductive polymer composite | 35 – 205 | 2000 | 3.17 | 0.124 |
| | Resonator 2 | | 35 – 205 | 1770 | 3.29 | 0.186 |
| | Resonator 1 | | 35 – 110 | 820 | 2.14 | 0.262 |
| S13 | Log-periodic dipole antenna (LPDA) and cavity filter | Ceramic | 20 – 1400 | 54.5 | 0.356 | 0.032 |
| S14 | 3D photonic crystal | Ceramic | 0 – 1400 | 84000 | 5 | 0.0714 |
| This work | Parallel plate capacitor and inductive coils | Conductive polymer composite | 20 – 110 | 6.8-7.0 | 0.028-0.036 | 0.38 – 0.55 |

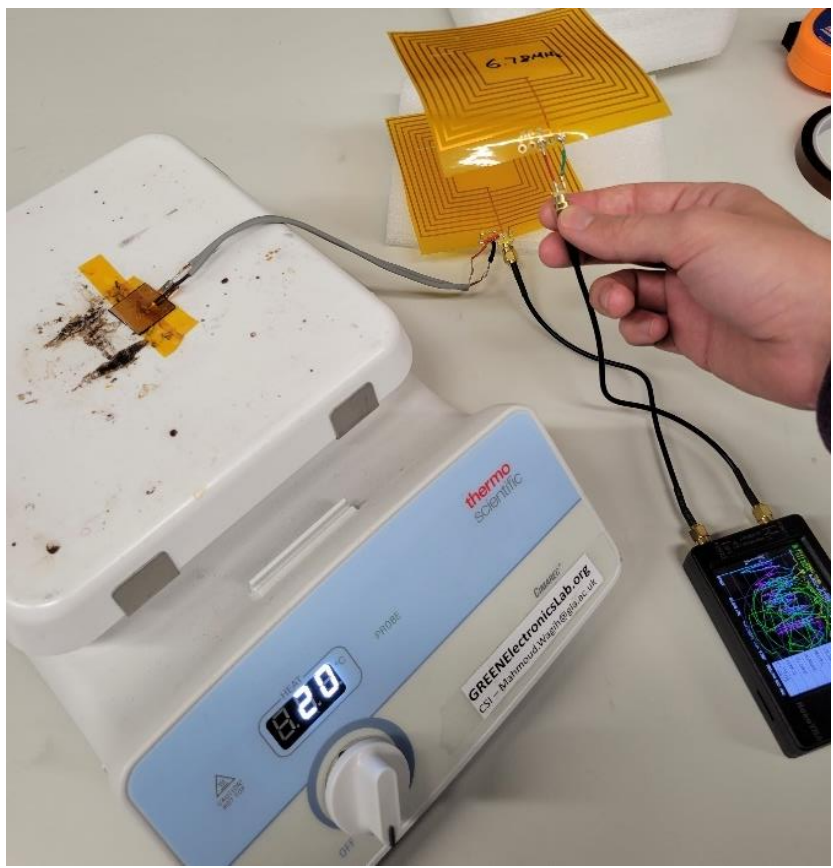

*Figure S7. Photograph of NanoVNA wireless temperature sensing setup.*

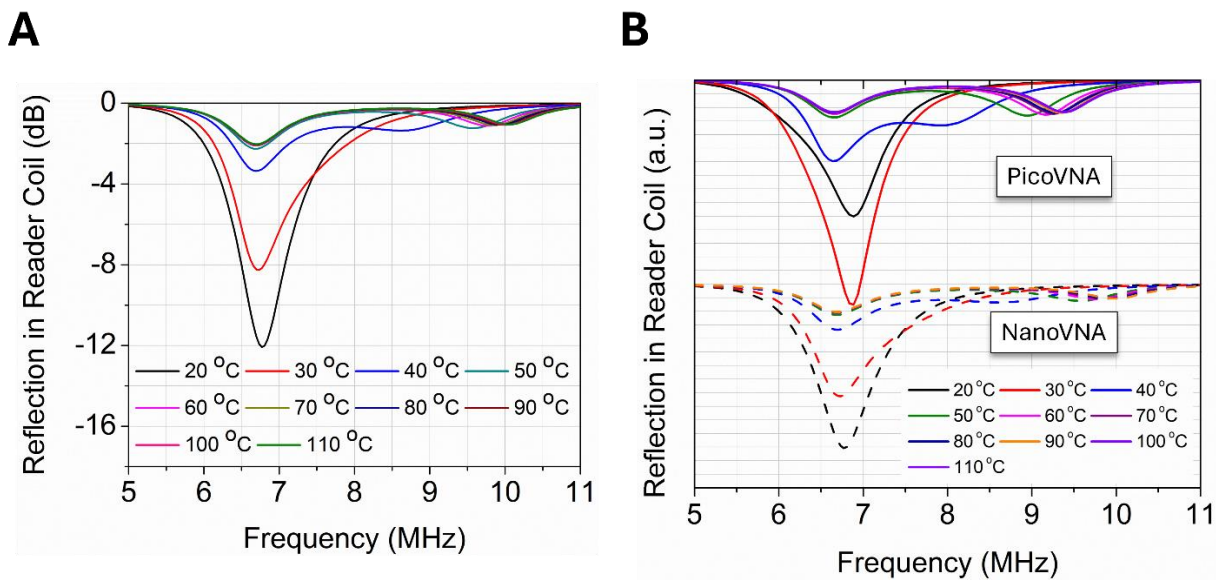

*Figure S8. Reflection data of 4 cm² wireless capacitor read by handheld (Nano) VNA and comparison between laboratory (Pico) VNA and hand-held (Nano) VNA. A) Reflection coefficient (S11) data from 20°C - 110°C for PDMS-CF measured with the NanoVNA and B) comparison of unscaled reflection coefficient (S11) for PicoVNA (solid line) and the portable/low-cost NanoVNA (dashed line).*

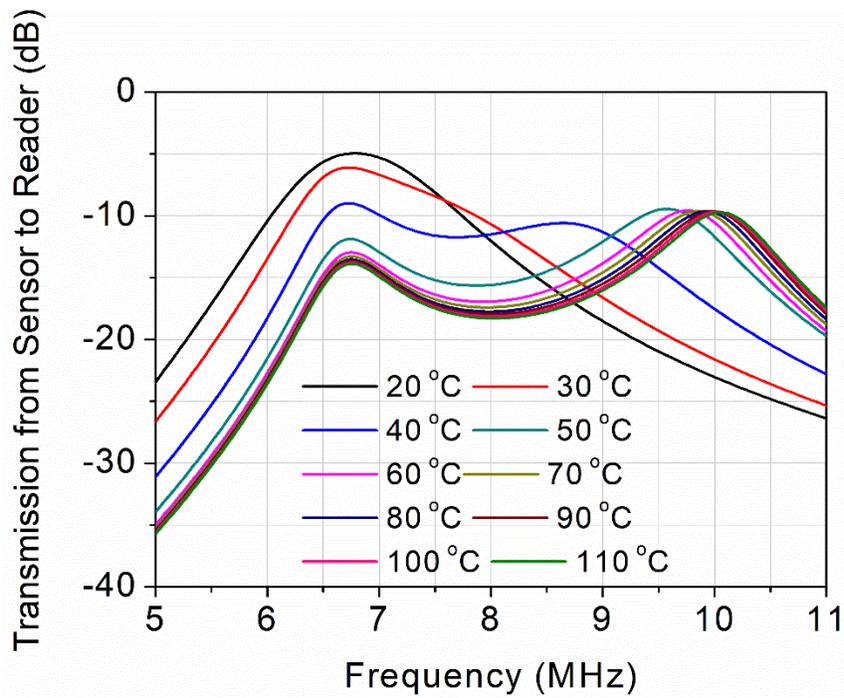

Figure S9. Transmission coefficient (S21) data from 20°C - 110°C for PDMS-CF measured with the NanoVNA.

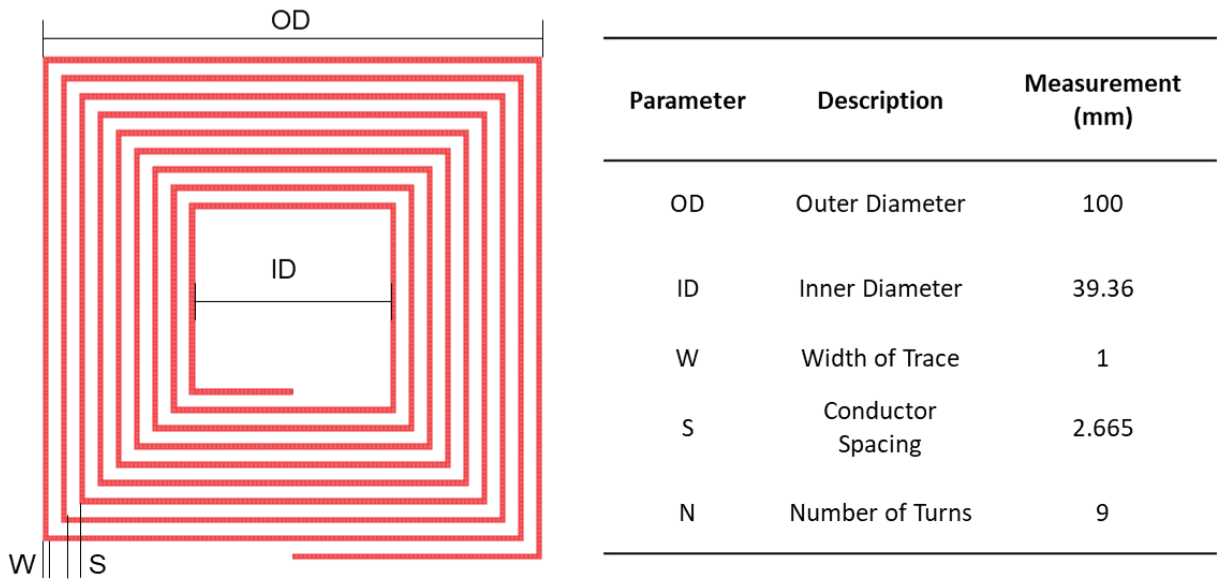

| Parameter | Description | Measurement (mm) |
|-----------|-------------|------------------|
| OD | Outer Diameter | 100 |
| ID | Inner Diameter | 39.36 |
| W | Width of Trace | 1 |
| S | Conductor Spacing | 2.665 |
| N | Number of Turns | 9 |

Figure S10. Layout and dimensions of the inductive coils used in this work.

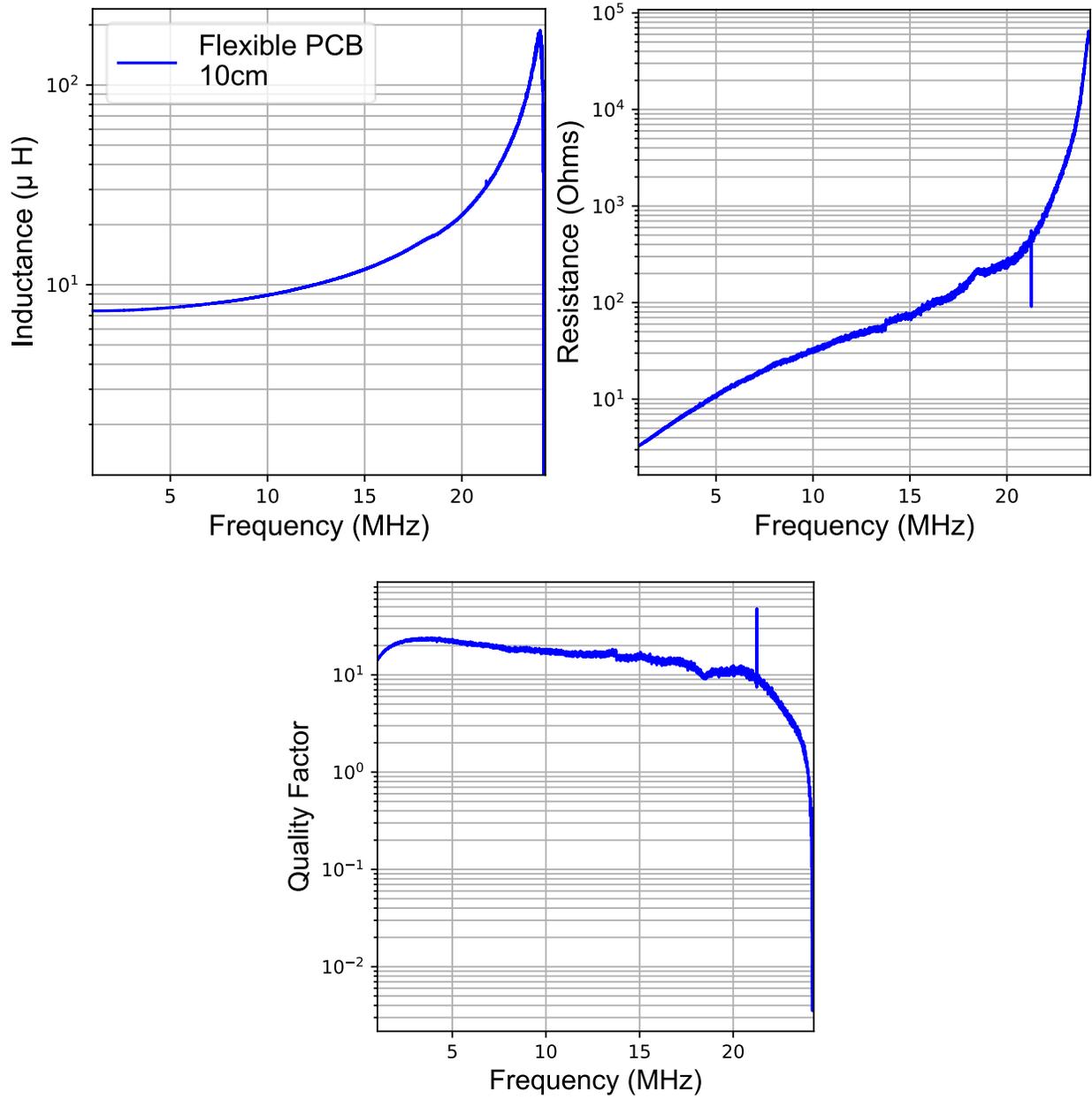

*FigureS11. Measured Inductance, Resistance, and Quality Factor of the coils used in the wireless chipless sensor interface. The reader coil was tuned using standard high-Q ceramic capacitors.*

(S1)   Kubina, B.; Schusler, M.; Mandel, C.; Mehmood, A.; Jakoby, R. Wireless High-Temperature Sensing with a Chipless Tag Based on a Dielectric Resonator Antenna. *Proc. IEEE Sensors* **2013**, p. 1-4. DOI: 10.1109/ICSENS.2013.6688181.

(S2)   Lu, F.; Tan, Q.; Ji, Y.; Guo, Q.; Guo, Y.; Xiong, J. A Novel Metamaterial Inspired High-Temperature Microwave Sensor in Harsh Environments. *Sensors* **2018**, *18* (9), 2879. DOI: 10.3390/s18092879.

(S3)   Tan, Q.; Ren, Z.; Cai, T.; Li, C.; Zheng, T.; Li, S.; Xiong, J. Wireless Passive Temperature Sensor Realized on Multilayer HTCC Tapes for Harsh Environment. *J. Sensors* **2015**, *2015, 124058*. DOI: 10.1155/2015/124058.

(S4)   Karim, H.; Delfin, D.; Chavez, L. A.; Delfin, L.; Martinez, R.; Avila, J.; Rodriguez, C.; Rumpf, R. C.; Love, N.; Lin, Y. Metamaterial Based Passive Wireless Temperature Sensor. *Adv. Eng. Mater.* **2017**, *19* (5), 1600741. DOI: 10.1002/adem.201600741.

(S5)   Wang, C.; Chen, L.; Tian, B.; Jiang, Z. High-Linearity Wireless Passive Temperature Sensor Based on Metamaterial Structure with Rotation-Insensitive Distance-Based Warning Ability. *Nanomaterials* **2023**, *13* (17), 2482. DOI: 10.3390/nano13172482.

(S6)   Tchafa, F. M.; Huang, H. Microstrip Patch Antenna for Simultaneous Temperature Sensing and Superstrate Characterization. *Smart Mater. Struct.* **2019**, *28* (10), 065019. DOI: 10.1088/1361-665X/ab2213.

(S7)   Leier, B.; Baghelani, M.; Iyer, A. K. A Microwave Stripline Ring Resonator Sensor Exploiting the Thermal Coefficient of Dielectric Constant for High-Temperature Sensing. *IEEE Sens. J.* **2022**, *22* (22), 21666–21675. DOI: 10.1109/JSEN.2022.3210779.

(S8)   Lin, J. H.; Kao, Y. H. Wireless Temperature Sensing Using a Passive RFID Tag with Film Bulk Acoustic Resonator. *Proc. - IEEE Ultrason. Symp.* **2008**, 2209–2212. DOI: 10.1109/ULTSYM.2008.0547.

(S9)   Idhaiam, K. S. V.; Pozo, P. D.; Sabolsky, K.; Sabolsky, E. M.; Sierros, K. A.; Reynolds, D. S. All-Ceramic LC Resonator for Chipless Temperature Sensing within High Temperature Systems. *IEEE Sens. J.* **2021**, *21* (18), 19771–19779. DOI: 10.1109/JSEN.2021.3094406.

(S10)  Cheng, H.; Ren, X.; Ebadi, S.; Chen, Y.; An, L.; Gong, X. Wireless Passive Temperature Sensors Using Integrated Cylindrical Resonator/Antenna for Harsh-Environment Applications. *IEEE Sens. J.* **2015**, *15* (3), 1453–1462. DOI: 10.1109/JSEN.2014.2363426.

(S11)  Ghafourian, M.; Bridges, G. E.; Nezhad, A. Z.; Thomson, D. J. Wireless Overhead Line Temperature Sensor Based on RF Cavity Resonance. *Smart Mater. Struct.* **2013**, *22* (7), 075010. DOI: 10.1088/0964-1726/22/7/075010.

(S12)  Wagih, M.; Shi, J.; Li, M.; Komolafe, A.; Whittaker, T.; Schneider, J.; Kumar, S.; Whittow, W.; Beeby, S. Wide-Range Soft Anisotropic Thermistor with a Direct

Wireless Radio Frequency Interface. *Nat. Commun.* **2024**, *15* (1), 452. DOI: 10.1038/s41467-024-44735-z.

(S13)  Peng, H.; Yang, X.; Wu, X.; Peng, W. A Wireless Temperature Sensor Applied to Monitor and Measure the High Temperature of Industrial Devices. *IEEE Trans. Antennas Propag.* **2024**, *72* (6), 5273–5282. DOI: 10.1109/TAP.2024.3389213.

(S14)  Sánchez-pastor, J.; Jakoby, R.; Benson, N.; Jiménez-sáez, A.; Benson, N.; Jim, A. A wireless W-band 3D-printed temperature sensor operating beyond 1000 °C. *Preprint* **2023**. DOI: 10.21203/rs.3.rs-3394921/v1.